\documentclass[12pt]{article}
\pdfoutput=1

\usepackage[utf8]{inputenc}
\usepackage[left=2.55cm, right=2.55cm, top=2.55cm, bottom=2.55cm]{geometry}
\usepackage{amsmath,amssymb,amsbsy,mathtools}
\usepackage{slashed}
\usepackage{xcolor}
\usepackage{graphicx}
\usepackage{url}
\usepackage{cancel}
\usepackage{cite}
\usepackage[colorlinks=true,allcolors=blue,pdfborder={0 0 0},linktocpage=false,pdfencoding=auto]{hyperref}
\usepackage{tabularx,booktabs}
\usepackage{multicol}
\usepackage{feynmp}
\usepackage{units}
\usepackage{xspace}
\usepackage[labelfont=bf]{caption}
\usepackage[section]{placeins}
\usepackage{subcaption}
\usepackage{soul}
\usepackage{bbold}
\usepackage{parskip}
\usepackage{float}
\usepackage{tabulary}
\usepackage{empheq}
\usepackage{tabu}

\definecolor{darkred}{rgb}{0.6,0,0}
\definecolor{darkpurple}{rgb}{0.5,0,0.5}

 \newcommand{\code}[1]{\texttt{#1}}
\newcommand{\beqn}{\begin{eqnarray}}
\newcommand{\eeqn}{\end{eqnarray}}
\def\non{\nonumber}

\def\non{\nonumber\\}

\begin{document}
\author{Amin Aboubrahim$^{a}$\footnote{\href{mailto:abouibra@union.edu}{abouibra@union.edu}}~~and Pran Nath$^b$\footnote{\href{mailto:p.nath@northeastern.edu}{p.nath@northeastern.edu}}
 \\~\\
$^{a}$\textit{\normalsize Department of Physics and Astronomy, Union College,} \\
\textit{\normalsize 807 Union Street, Schenectady, NY 12308, U.S.A.} \\
$^{b}$\textit{\normalsize Department of Physics, Northeastern University,} \\
\textit{\normalsize 111 Forsyth Street, Boston, MA 02115-5000, U.S.A.} \\
}

\title{\vspace{-2cm}
\vspace{1cm}
\large \bf {Interacting ultralight dark matter and dark energy and fits to cosmological data in a field theory approach}
 \vspace{0.5cm}}
\date{}
\maketitle

\vspace{0.5cm}

\begin{abstract}

The description of dark matter as a pressure-less fluid and of dark energy as a cosmological constant, both minimally coupled to gravity, constitutes the basis of the concordance 
$\Lambda$CDM model. However, the concordance model is based on using equations of motion directly for the fluids with constraints placed on their sources, and
lacks an underlying Lagrangian. In this work, we propose a Lagrangian model of two spin zero  fields describing dark energy and dark matter with an interaction term between the two along with self-interactions. We study the background evolution of the fields as well as their linear perturbations, suggesting an alternative to $\Lambda$CDM with dark matter and dark energy being fundamental dynamical fields. The parameters of the model are extracted using a Bayesian inference tool based on multiple cosmological data sets which include those of Planck (with lensing), BAO, Pantheon, SH0ES, and WiggleZ. Using these data, we set constraints on the dark matter mass and the interaction strengths. Furthermore, we find that the model is able to alleviate the Hubble tension for some data sets while also resolving the $S_8$ tension. 

\end{abstract}

\numberwithin{equation}{section}

\newpage

{  \hrule height 0.4mm \hypersetup{colorlinks=black,linktocpage=true} \tableofcontents
\vspace{0.5cm}
 \hrule height 0.4mm}

\section{Introduction} \label{sec:int}

Analyses of the most recent data from the Planck satellite experiment~\cite{Aghanim:2018eyx} indicate that the composition of the universe consists of roughly 5\%  visible matter, about 25\%  dark matter (DM) and the rest, about 70\% dark energy (DE). For it to drive the accelerated expansion of the universe, dark energy is characterized by negative pressure $p_{\rm de}$ so that its equation of state $w_{\rm de}= p_{\rm de}/\rho_{\rm de}$, where $\rho_{\rm de}$ is the energy density of dark energy, is negative with $w_{\rm de}<-1/3$. Good fits to the experimental data can be obtained with $w_{\rm de}=-1$ along with a pressureless cold dark matter with $w_{\rm dm}=0$. The two fluids constitute the basis of the concordance $\Lambda$CDM model which is known as the Standard Model of Cosmology. However, the theoretical origin of dark energy remains unclear and there are a variety of models to explain the origin of dark energy~\cite{Fujii:1982ms,Ford:1987de,Wetterich:1987fm,Ratra:1987rm,Chiba:1997ej,Ferreira:1997au,Ferreira:1997hj,Copeland:1997et,Caldwell:1997ii,Zlatev:1998tr,Chiba:1999ka,Armendariz-Picon:2000nqq,Armendariz-Picon:2000ulo,Kamenshchik:2001cp,delaMacorra:1999ff,Ng:2001hs,Corasaniti:2002vg,Caldwell:2005tm,Scherrer:2007pu}. Most of these belong to a generic class known as quintessence models. A quintessence field is a form of an ultralight axion (ULA) whose mass is $\mathcal{O}(10^{-33})$ eV which is $\sim H_0$ (today's Hubble constant). Quintessence models are often categorized as ``thawing" or ``freezing"~\cite{Caldwell:2005tm,Pantazis:2016nky} depending on how they evolve with time driven by their axionic potential. Quintessence is capable of fully providing an explanation of the accelerated expansion of the universe without having to invoke a cosmological constant. Further, dark energy as quintessence is a dynamical axionic field which can be useful in explaining the coincidence problem~\cite{Armendariz-Picon:2000nqq,Armendariz-Picon:2000ulo}. There are also models which attempt to explain dark energy within a modified gravity framework.  For a detailed exposition of them, the reader is directed to reviews on the subject, such as refs.~\cite{Tsujikawa:2013fta,Carroll:2000fy,Peebles:2002gy,Copeland:2006wr}.

ULAs can also be viable dark matter candidates if they have a mass $\mathcal{O}(10^{-22})$ eV~\cite{Hu:2000ke,Ringwald:2012hr,Kim:2015yna,Hui:2016ltb,Halverson:2017deq}. Given their very small mass, these DM particles have de Broglie wavelengths the size of galaxies. On large scales, ULAs mimic cold dark matter (CDM) and above a certain mass, they become indistinguishable from CDM. However, on small scales, ULAs suppress structure formation owing to quantum pressure from Heisenberg uncertainty principle. This property has made ULA dark matter (also known as fuzzy DM~\cite{Hu:2000ke} when it comprises the entire DM relic density), a potential candidate providing a solution to the core-cusp problem~\cite{deBlok:2009sp} \footnote{
{We note here in passing that core-cusp problem and other short distance galaxy anomalies can
also be explained more generally by self interacting dark matter~see \cite{Spergel:1999mh,Aboubrahim:2020lnr} and the papers referenced therein.}}.
 For a typical scalar field DM $\chi$ with a potential $V(\chi)=(1/2) m_\chi^2 \chi^2$, the field slowly approaches the minimum of its potential as the universe expands. Once the time scale set by $m^{-1}$ becomes much smaller than the Hubble time $H^{-1}$, the field starts oscillating around the minimum of its potential~\cite{Turner:1983he}. During oscillation, the DM energy density redshifts as $a^{-3}$, where $a$ is the scale factor. This means that the scalar field now behaves as CDM diluting away as a pressureless matter field with $w_\chi=0$. These oscillations are also present at the level of linear perturbations where an oscillating field admits an effective sound speed that remains appreciable at small scales. This pressure support coming from the sound speed stalls the growth of perturbation, causing the erasure of structure at small scales which is also reflected as a cut-off in the matter power spectrum. Note that not all perturbation modes experience suppression of growth, only those below the Jeans scale~\cite{Hlozek:2014lca}.

The quadratic scalar potential for DM mentioned above has been studied extensively in the literature~\cite{Sahni:1999qe,Matos:2000ng,Amendola:2005ad,Matos:2008ag,Hwang:2009js,Marsh:2010wq,Glennon:2023jsp,Park:2012ru,Li:2013nal,Urena-Lopez:2015gur} (see refs.~\cite{Marsh:2015xka,Magana:2012xe} for reviews) along with anharmonic corrections resulting from the addition of the self-interaction quartic term~\cite{Cembranos:2018ulm,Foidl:2022bpn}. Sizable DM self-interactions can leave imprints on the CMB power spectrum as well as the matter power spectrum and therefore can be constrained by observations. Ref.~\cite{Foidl:2022bpn} discussed the case of a complex ultralight scalar field  with strong repulsive self-interaction (SI) and compared it to the cases of a complex scalar field with no SI and a ULA with no SI.  The case of a complex scalar field with strong SI starts with a phase of stiff matter domination with $w_\chi=1$ in the early universe followed by a transition to a radiation-like phase, $w_\chi=1/3$, before entering a CDM-like phase with $w_\chi=0$ right before radiation-matter equality. On the other hand, complex scalar fields with no SI do not experience a radiation-like phase and the transition from $w_\chi=1$ to $w_\chi=0$ happens almost suddenly much before radiation-matter equality. As for ULAs, the evolution starts with the field frozen in place due to Hubble friction. In other words, the field's kinetic energy is much smaller than its potential energy which makes the ULA behave as dark energy in the early universe, commonly dubbed as early dark energy (EDE)~\cite{Doran:2006kp,Agrawal:2019lmo} with $w_\chi=-1$. In the presence of strong SI, ULAs also experience an intermediate radiation-like phase before the field starts its coherent oscillations around the minimum of the potential. For ULAs lighter than $\sim 10^{-22}$ eV, CMB observations using the Planck data indicate that ULAs cannot make up the entire dark matter relic density, but rather only a fraction $f_\chi=\Omega_\chi(z_c)/\Omega_{\rm tot}(z_c)$, where $z_c$ is the redshift at which the field becomes dynamical. It was shown in ref.~\cite{Poulin:2018dzj} that for $10\lesssim 1+z_c\lesssim 3\times 10^4$, $f_\chi\lesssim 0.004$ for a potential of the form $V(\chi)\propto 1-\cos(\chi/F)$. The constraint relaxes for increasing $z_c$. In ref.~\cite{Hlozek:2014lca}, the authors show that a ULA in the mass range $10^{-32}\,\text{eV}\leq m_\chi\leq 10^{-25.5}\,\text{eV}$ is bound by the constraint $\Omega_\chi h^2\leq 0.006$ at 95\% CL. For masses greater than $10^{-24}$ eV, a ULA is indistinguishable from the standard CDM at linear scales.   

More recently there have been several models of dark energy interacting with dark matter under various assumptions on the couplings~\cite{Pourtsidou:2013nha,Pourtsidou:2016ico,Linton:2017ged,Chamings:2019kcl,Pan:2019gop,Bonici:2018qli,Yang:2019uzo,Pan:2020zza} (for a review see~\cite{Wang:2016lxa,Wang:2024vmw,Bamba:2012cp}). One of the ways such a coupling can be introduced is at the level of an interaction Lagrangian using a variational approach~\cite{Boehmer:2015kta,Boehmer:2015sha,Archidiacono:2022iuu,Rezazadeh:2022lsf} or at the level of energy density continuity equations~\cite{Potter:2011nv,DiValentino:2019jae,DiValentino:2019ffd,Escamilla:2023shf,Bernui:2023byc,Sharma:2021ayk,Sharma:2021ivo} where both DM and DE are fluids~\cite{Yang:2022csz}, DM is a fluid while DE is quintessence~\cite{Amendola:1999er,Kase:2019veo,Perez:2021cvg,Lee:2022cyh} and both DM and DE are scalar fields~\cite{Garcia-Arroyo:2024tqq,Beyer:2010mt,Beyer:2014uqa,vandeBruck:2022xbk}. Many of the interacting DM-DE models were invoked to try and explain the recent tensions in cosmology (for a review see ref.~\cite{Abdalla:2022yfr}). These tensions correspond to discrepancies between local measurements of observables~\cite{Riess:2019cxk,Wong:2019kwg} and model-dependent results from CMB data analysis at early times~\cite{Planck:2018nkj,ACT:2020gnv,SPT-3G:2021wgf}. One of the most severe of these tensions is the Hubble tension which corresponds to a disagreement, at the $5\sigma$ level, between local measurements from the SH0ES collaboration~\cite{Riess:2021jrx} using Cepheid-calibrated supernovae and early time predictions using the CMB data from the Planck collaboration~\cite{Planck:2018vyg}. For a recent analysis of improved Planck constraints on axion-like early dark energy to resolve Hubble tension\footnote{We note that the Hubble tension has also been addressed by numerous works based on thermal DM candidates and interacting sectors. See ref.~\cite{Aboubrahim:2022gjb} and the papers cited therein for discussions of models modifying early time physics. Note, however, that ref.~\cite{Vagnozzi:2023nrq} suggests that early time physics alone cannot resolve this tension as does uncoupled quintessence~\cite{Banerjee:2020xcn}.}, see ref.~\cite{Efstathiou:2023fbn}. Furthermore, measurements of the clustering strength of matter in the universe at the large-scale structure from weak gravitational lensing and galaxy clustering surveys~\cite{KiDS:2020suj,Joudaki:2019pmv,DES:2021wwk,Heymans:2020gsg,DES:2020ahh,Kazantzidis:2018rnb} have also shown to be inconsistent with predictions using the matter clustering power from the CMB anisotropies based on $\Lambda$CDM. This $2-3\sigma$ tension shows up in the parameter $S_8\equiv\sigma_8\sqrt{\Omega_{\rm m}/0.3}$ which is the weighted amplitude of the variance in matter fluctuations for spheres of size $8h^{-1}$Mpc. 

In this work we investigate a cosmological model based on a field theory approach where DM is a scalar field and DE an axionic field, both being ultralight interacting fields. In particular, DE is a quintessence field while DM is an ULA comprising the entire DM density. Our model presents a coupling between DM and DE originating from an interaction term in a Lagrangian which has not been realized or constrained in a cosmological model before. Furthermore, the DM and DE potential terms and the interaction term generate non-standard source terms in the DM and DE continuity equations, which require a redefinition of the total energy density and pressure of the DM and DE fields. We carry out the analysis by first deriving the background and linear perturbation equations of the coupled DM-DE fields and numerically solving them in order to extract constraints on the free parameters of the model using recent cosmological data sets.  

The outline of the rest of the paper is as follows: Section~\ref{sec:model} presents the details of our model which consists of a dark energy field with an axionic potential and a dark matter field with self-interaction as well as a DM-DE interaction term. In sections~\ref{sec:background} and~\ref{sec:perturb} we derive the background and linear perturbation equations of the fields and the technique used to average over fast oscillations is discussed in section~\ref{sec:average}. The numerical analysis is presented in section~\ref{sec:na} and conclusions in section~\ref{sec:conc}. {In Appendices~\ref{app:A} and~\ref{app:B} we give the perturbation equations in both the synchronous and newtonian gauges, before and after the onset of rapid oscillations. }

\section{Interacting dark energy and dark matter model\label{sec:model}}

The model of dark matter and dark energy we consider in this work is based on a particle physics Lagrangian of an axionic field $\phi$  denoting dark energy and a real scalar field $\chi$ representing dark matter. The action of the coupled $\phi$-$\chi$  system is given by 
\begin{align}
A= \int \text{d}^4 x \sqrt{-g} \left[-\frac{1}{2}\phi^{,\mu} \phi_{,\mu} - \frac{1}{2} \chi^{,\mu} \chi_{,\mu}   
       -V(\phi, \chi)\right],         
\end{align} 
where $g=\text{det}(g_{\mu\nu})$ is the determinant of the metric and  $V(\phi,\chi)$, the potential of the system, is taken to be of the form
\begin{equation}
    V(\phi,\chi)=  V_1(\chi)+ V_2(\phi) + V_3(\phi,\chi).
    \label{vpc}
\end{equation}
The DM-only potential $V_1(\chi)$ is given by
\begin{align}
 V_1(\chi)&=\frac{1}{2} m_\chi^2 \chi^2+ \frac{\lambda}{4} \chi^4, 
\label{v1}
\end{align}   
where we take the self-interaction term to be repulsive, i.e., $\lambda>0$. 
For DE we take the typical axionic potential
\begin{align}
V_2(\phi)&= \mu^4 \left[1+\cos\left(\frac{\phi}{F}\right)\right],
\label{v2}
\end{align} 
where $\mu$ has units of mass and $F$ is the axion decay constant and it is order of the Planck mass $M_{\rm Pl}$.  A potential of this type has been used in quintessence. For an overview, see, e.g., ref.~\cite{Tsujikawa:2013fta}. In the analysis, we also consider a phenomenological 
 interaction term between the two fields given by
\begin{align}
V_3(\phi,\chi)&= \frac{\tilde\lambda}{2}\chi^2\phi^2,            
\end{align} 
where the dimensionless parameter $\tilde\lambda$ is the strength of the DM-DE interaction. One can also include other interaction terms in the Lagrangian but this will be deferred to future work. Now with the potential $V(\phi,\chi)$ of Eq.~(\ref{vpc}), we can calculate the actual masses of $\chi$ and $\phi$ as 
\begin{align}
    M^2_\chi&=V_{,\chi\chi}\equiv\frac{\partial^2 V}{\partial\chi^2}=m^2_\chi+3\lambda\chi^2+\tilde{\lambda}\phi^2, \\
    M^2_\phi&=V_{,\phi\phi}\equiv\frac{\partial^2 V}{\partial\phi^2}=-\frac{\mu^4}{F^2}\cos\left(\frac{\phi}{F}\right)+\tilde{\lambda}\chi^2.
\end{align}

\section{Background equations \label{sec:background}}

Following the period of rapid inflation, the universe in our model becomes populated with the Standard Model (SM) particles: baryons, photons and neutrinos. Furthermore, it contains two ultralight fields: dark matter $\chi$ and dark energy $\phi$ whose Lagrangian was defined in the previous section. Similar to the assumption carried out in $\Lambda$CDM, we consider a flat, homogeneous and isotropic universe characterized by the Friedmann-Lema\^{i}tre-Roberston-Walker (FLRW) metric. The line element is
\begin{equation}
    \text{d}s^2=g_{\mu\nu}\text{d}x^\mu \text{d}x^\nu=a^2(-\text{d}\tau^2+\gamma_{ij}\text{d}x^i \text{d}x^j),
\end{equation}
where $a$ is the time-dependent scale factor, $\gamma_{ij}$ are the spatial components of the metric and $\tau$ is the conformal time which is related to the cosmic time by $\text{d}\tau=\text{d}t/a(t)$. An essential ingredient for studying cosmic evolution are the Einstein field equations
\begin{eqnarray}
    R_{\mu\nu}-\frac{1}{2}g_{\mu\nu}R=8\pi G T_{\mu\nu},
\end{eqnarray}
where $R_{\mu\nu}$ is the Ricci tensor, $R$ is the Ricci scalar, $G$ is Newton's gravitational constant and $T_{\mu\nu}$ is the stress-energy tensor. Remember that in our model there is no cosmological constant. The axionic field $\phi$ takes up that role and is contained inside $T_{\mu\nu}$ along with the DM field $\chi$ and the rest of the SM particles. The $00$ component of the Einstein equation gives us the Friedmann equation relating the Hubble parameter $H=\dot{a}/a$ (the dot represents a derivative with respect to cosmic time $t$) or $\mathcal{H}=a^\prime/a=aH$ (a prime corresponds to a derivative with respect to conformal time) so that
\begin{equation}
    \mathcal{H}^2(\tau)=\frac{a^2}{3 m_{\rm Pl}^2}\sum [\rho_b(\tau)+\rho_r(\tau)+\rho_D(\tau)],
\end{equation}
where $\rho_b$, $\rho_r$, $\rho_D$ are the energy densities of baryons, radiation, and the total 
energy density of dark matter and dark energy, and $m_{\rm Pl}=1/\sqrt{8\pi G}$ is the reduced Planck mass. The quantity $\rho_D(\tau)$ is given by $\rho_\chi(\tau)+\rho_\phi(\tau) -V_3(\tau)$. As seen below $V_3(\tau)$ appears in the expressions for both $\rho_\chi(\tau)$ and $\rho_\phi(\tau)$
and $-V_3(\tau)$ in the expression for $\rho_D$ is to eliminate double counting.

Thus for the energy densities of the background fields we write
\begin{align}
    \label{rhop}
    \rho_\phi=\frac{1}{2a^2}\phi_0^{\prime 2}+\bar{V}_2(\phi)+\bar{V}_3(\phi,\chi), \\
    \rho_\chi=\frac{1}{2a^2}\chi_0^{\prime 2}+\bar{V}_1(\chi)+\bar{V}_3(\phi,\chi), 
    \label{rhoc}
\end{align}
where $\chi_0$ and $\phi_0$ denote the background fields and a bar over the potential terms indicates that they are a function of the background fields. Similarly, using the $ij$ components of the stress-energy tensor, one can derive the pressure of the DM and DE fields
\begin{align}
    p_\phi=\frac{1}{2a^2}\phi_0^{\prime 2}-\bar{V}_2(\phi)-\bar{V}_3(\phi,\chi), \\
    p_\chi=\frac{1}{2a^2}\chi_0^{\prime 2}-\bar{V}_1(\chi)-\bar{V}_3(\phi,\chi).
    \label{pp}
\end{align}
In order to calculate the energy densities and pressure of our DM and DE fields, we need to track the evolution of the fields themselves. This is done via the Klein-Gordon (KG) equation, which, for the field $\chi$, is calculated using the equation of motion
\begin{equation}
    0=\frac{1}{\sqrt{-g}}\partial_\mu \Big(\sqrt{-g}\, g^{\mu\nu}\partial_\nu \chi\Big)-V_{,\chi}\,,
    \label{eqmotion}
\end{equation}
where $V_{,\chi}\equiv \partial V/\partial\chi$. The resulting KG equations of DM and DE are
\begin{align}
\label{KGc0}
&\chi_0^{\prime\prime}+2\mathcal{H}\chi_0^\prime+a^2(\Bar{V}_1+\Bar{V}_3)_{,\chi}=0, \\
&\phi_0^{\prime\prime}+2\mathcal{H}\phi_0^\prime+a^2(\Bar{V}_2+\Bar{V}_3)_{,\phi}=0,
\label{KGp0}
\end{align}
where $\bar V(\phi, \chi)\equiv V(\phi_0, \chi_0)$ and $\bar V_{1,\chi} \equiv (V_{1,\chi})_{\chi=\chi_0}$, etc. 

Using the KG equations with the energy density and pressure equations, we arrive at the continuity equations
\begin{align}
\label{rhocont1}
&\rho^\prime_\phi+3\mathcal{H}(1+w_\phi)\rho_\phi=Q_\phi\,, \\
&\rho^\prime_\chi+3\mathcal{H}(1+w_\chi)\rho_\chi=Q_\chi\,.
\label{rhocont2}
\end{align}
The source terms $Q_\phi=\Bar{V}_{3,\chi}\chi_0^\prime$ and $Q_\chi=\Bar{V}_{3,\phi}\phi_0^\prime$ represent the couplings between the two fields. These terms have a well-defined particle physics origin and as a consequence appear naturally in the continuity equations rather than being ad hoc terms. The equations of state of the two fields are defined as $w_i=p_i/\rho_i$.
In the analysis here, we take into account interactions between the fields $\phi$ and $\chi$
in the dark sector, but ignore the possible feeble interactions between the dark sector and the
visible sector. In this case the conservation of total energy density  in the dark sector is given by
\begin{equation}
    \rho^\prime+3\mathcal{H}(\rho+p)=0,
\label{continuity}
\end{equation}
with $p$ being the total pressure and we have dropped the subscript $D$ on $\rho$ and $p$
since the analysis is focused on the dark sector. 
We note here that $Q_\phi$ and $Q_\chi$ that appear in Eq.~(\ref{rhocont2}) do not 
satisfy the relation $Q_\phi=-Q_\chi$ which has been used in numerous dark matter-dark energy
analyses. In fact the constraint $Q_\phi=-Q_\chi$ cannot arise in any consistent 
 Lagrangian theory. In a Lagrangian field theory energy conservation equation Eq.~(\ref{continuity})
 is automatic without the necessity of any additional constraints.

\section{Linear perturbations\label{sec:perturb}}

We discussed in the previous section the evolution of the background equations which assumes a homogeneous universe, i.e., the fields only depend on time. However, our universe is clearly not homogeneous and the fields have both time and position dependence, i.e., $\chi(t,\Vec{x})=\chi_0(t)+\chi_1(t,\Vec{x})+\cdots$ and $\phi(t,\Vec{x})=\phi_0(t)+\phi_1(t,\Vec{x})+\cdots$, where $\chi_1(t,\Vec{x})$ and $\phi_1(t,\Vec{x})$ are first order perturbations of the fields. Remarkably, deviations from the background field are small in the early universe and one can use linear perturbation theory to describe the growth of structure in the universe. Non-linear growth becomes important in the late universe and at small scales and is beyond the scope of this work.  
 Thus in the analysis here we will consider only linear effects.

We start by perturbing the metric around its background value: $g^{\mu\nu}=\bar{g}^{\mu\nu}+\delta g^{\mu\nu}$, so that in the general gauge~\cite{Hu:2003hjx}, we have 
\begin{equation}
\label{gen}
\begin{cases}
    g^{00}=-a^{-2}(1-2A), & \\
    g^{0i}=-a^{-2}B^i, & \\
    g^{ij}=a^{-2}(\gamma^{ij}-2H_L \gamma^{ij}-2H_T^{ij}),
\end{cases}
\end{equation}
where $A$ is a scalar potential, $B^i$ a vector shift, $H_L$ a scalar perturbation to the spatial curvature and $H_T^{ij}$ a trace-free distortion to the spatial metric. In the literature, the gauges of choice are mainly the synchronous gauge and the conformal (Newtonian) gauge. In the synchronous gauge, the components $g^{00}$ and $g^{0i}$ are not perturbed and so the line element has the form: $\text{d}s^2=a^2(\tau)\left[-\text{d}\tau^2+(\delta_{ij}+h_{ij})\text{d}x^i \text{d}x^j\right]$. Therefore one has 
\begin{align}
    &A=B=0, \nonumber \\
    &H_L=\frac{1}{6}h,
    \label{sync}
\end{align}
where $h$ represents the trace of the metric perturbations $h_{ij}$. This gauge is easy to use especially in numerical codes but has some disadvantages, one of which is that it does not completely fix the gauge degrees of freedom. This issue is overcome in $\Lambda$CDM due to the presence of a pressureless fluid (CDM) which is the extra ingredient required to fix the gauge. However, this remedy is spoiled when considering a light scalar field as DM\footnote{We can still use the synchronous gauge in this work and we will come back to this issue in the numerical analysis part.}. On the other hand, the conformal gauge~\cite{Mukhanov:1990me} leaves no ambiguities and can easily accommodate a scalar field as DM. This gauge is characterized by the choice
\begin{align}
    &B=H_T=0,  \nonumber \\
    &A\equiv \Psi ~~~~~ (\text{Newtonian potential}), \nonumber \\
    &H_L \equiv \Phi ~~~ (\text{Newtonian curvature}).
    \label{conf}
\end{align}
We will carry out our calculations in the general gauge and then present our final results in both the synchronous and conformal gauges based on the above recipe. 

We now turn our attention to the stress-energy tensor. The perturbed object is $T^{\mu}_{\nu}=\bar{T}^{\mu}_{\nu}+\delta T^{\mu}_{\nu}$, so that
\begin{align}
    T^{0}_0&=-\rho-\delta\rho \nonumber \\
    T^{0}_i&=(\rho+p)(v_i-B_i) \nonumber \\
    T^{i}_0&=-(\rho+p)v_i \nonumber \\
    T^{i}_j&= (p+\delta p)\delta^i_j+p\Pi^i_j , 
\end{align}
with $\Pi^i_j$ representing the anisotropic stress, $v_i$ the 3-velocity,  $\delta\rho$ and $\delta p$ being the density and pressure perturbations, respectively. It immediately follows that the density and pressure perturbations of the two fields, $\chi$ and $\phi$, in the general gauge are given by
\begin{align}
    \label{drhop}
    \delta\rho_\phi&=\frac{1}{a^2}\phi_0^\prime\phi_1^\prime-\frac{1}{a^2}\phi_0^{\prime 2}A+(\Bar{V}_2+\Bar{V}_3)_{,\phi}\phi_1+\Bar{V}_{3,\chi}\chi_1, \\
    \delta p_\phi&=\frac{1}{a^2}\phi_0^\prime\phi_1^\prime-\frac{1}{a^2}\phi_0^{\prime 2}A-(\Bar{V}_2+\Bar{V}_3)_{,\phi}\phi_1-\Bar{V}_{3,\chi}\chi_1, \\
    \delta\rho_\chi&=\frac{1}{a^2}\chi_0^\prime\chi_1^\prime-\frac{1}{a^2}\chi_0^{\prime 2}A+(\Bar{V}_1+\Bar{V}_3)_{,\chi}\chi_1+\Bar{V}_{3,\phi}\phi_1, \\
    \delta p_\chi&=\frac{1}{a^2}\chi_0^\prime\chi_1^\prime-\frac{1}{a^2}\chi_0^{\prime 2}A-(\Bar{V}_1+\Bar{V}_3)_{,\chi}\chi_1-\Bar{V}_{3,\phi}\phi_1.
    \label{dpchi}
\end{align} 
From the perturbed stress-energy tensor, we have the off-diagonal term $\delta T^0_i=-a^{-2}\phi_0^\prime \delta\phi_{,i}$. Taking the spatial derivative and switching to Fourier space, we obtain the velocity divergence $\theta=ik^i v_i$ of the fields
\begin{align}
    (\rho_\phi+p_\phi)\theta_\phi&=\frac{k^2}{a^2}\phi_0^\prime\phi_1, \\
    (\rho_\chi+p_\chi)\theta_\chi&=\frac{k^2}{a^2}\chi_0^\prime\chi_1\,.
\end{align}
In many cases, the use of $\theta$ may cause some numerical instabilities. To circumvent this issue, we define $\Theta_i\equiv (1+w_i)\theta_i$ so that
\begin{align}
\label{uphi}
    \rho_\phi \Theta_\phi=\frac{k}{a^2}\phi_0^\prime\phi_1, \\
    \rho_\chi \Theta_\chi=\frac{k}{a^2}\chi_0^\prime\chi_1\,.
\label{cphi}    
\end{align}
Using Eq.~(\ref{eqmotion}) and picking only the first order perturbations, we arrive at the Klein-Gordon equations for the perturbations of the two fields in the general gauge
\begin{align}
\label{KGp1}
&\phi_1^{\prime\prime}+2\mathcal{H}\phi_1^\prime+(k^2+a^2\Bar{V}_{,\phi\phi})\phi_1+a^2\Bar{V}_{,\phi\chi}\chi_1+2a^2\Bar{V}_{,\phi}A+(3H_L^\prime-A^\prime+kB)\phi_0^\prime=0, \\
&\chi_1^{\prime\prime}+2\mathcal{H}\chi_1^\prime+(k^2+a^2\Bar{V}_{,\chi\chi})\chi_1+a^2\Bar{V}_{,\chi\phi}\phi_1+2a^2\Bar{V}_{,\chi}A+(3H_L^\prime-A^\prime+kB)\chi_0^\prime=0.
\label{KGc1}
\end{align}
In principle all the tools needed to calculate the density and velocity perturbations are in place. The background fields $\chi_0$ and $\phi_0$ and their perturbations $\chi_1$ and $\phi_1$ are calculated by solving the Klein-Gordon equations, Eqs.~(\ref{KGc0}), (\ref{KGp0}), (\ref{KGp1}) and~(\ref{KGc1}). Then the density and pressure perturbations are evaluated using Eqs.~(\ref{drhop})$-$(\ref{dpchi}). Finally, we calculate the density contrast for the fields which is given by
\begin{equation}
    \delta_i\equiv\frac{\delta\rho_i}{\bar{\rho}_i}=\frac{\rho_i(t,\Vec{x})-\bar{\rho}_i(t)}{\bar{\rho}_i},
\end{equation}
as well as the velocity divergence of the fields from Eqs.~(\ref{uphi}) and~(\ref{cphi}). Solving the KG equations can be computationally demanding especially when the DM field starts its rapid oscillations when $M_{\chi}^{-1}\ll H^{-1}$. For this reason, it is more practical to turn these equations into differential equations in $\delta_i$ and $\Theta_i$~\cite{Turner:1983he} (known as the fluid equations) using the generalized dark matter scheme~\cite{Hu:1998kj}. This scheme requires the field equation of state $w_i$ and a time and scale-dependent sound speed $c_{s}^2$~\cite{Hu:2000ke,Hwang:2009js,Noh:2017sdj}
\begin{equation}
    c_s^2=\frac{\delta p}{\delta \rho}.
\end{equation}
For the DM field $\chi$, we obtain a first order differential equation of the density contrast
\begin{align}
\delta_\chi^\prime&=\left[3\mathcal{H}(w_\chi-c_{s\chi}^2)-\frac{Q_\chi}{\rho_\chi}\right]\delta_\chi+\frac{3\mathcal{H}Q_\chi}{\rho_\chi(1+w_\chi)}(c_{s\chi}^2-c^2_{\chi_{\rm ad}})\frac{\Theta_\chi}{k}-9\mathcal{H}^2(c_{s\chi}^2-c^2_{\chi_{\rm ad}})\frac{\Theta_\chi}{k}-\Theta_\chi k \nonumber \\
&+\frac{a^2}{k}\frac{\rho_\phi}{\rho_\chi}\Bar{V}_{3,\phi\phi}\Theta_\phi+\frac{1}{\rho_\chi}\Bar{V}_{3,\chi\phi}\phi_0^\prime\chi_1+\frac{1}{\rho_\chi}\bar{V}_{3,\phi}\phi_1^\prime-(3H_L^\prime+kB)(1+w_\chi),
\label{dchi}
\end{align}
and for the velocity divergence
\begin{align}
\Theta^\prime_\chi&=(3c_{s\chi}^2-1)\mathcal{H}\Theta_\chi+k\delta_\chi c_{s\chi}^2+3\mathcal{H}(w_\chi-c^2_{\chi_{\rm ad}})\Theta_\chi \non
&~~~-\frac{Q_\chi}{\rho_\chi}\left(1+\frac{c_{s\chi}^2-c^2_{\chi_{\rm ad}}}{1+w_\chi}\right)\Theta_\chi+\frac{k}{\rho_\chi}\Bar{V}_{3,\phi}\phi_1+k(1+w_\chi)A.
\label{tchi}
\end{align}
In the above expressions, we have introduced the adiabatic sound speed $c^2_{\chi_{\rm ad}}$ which is a quantity that depends only on background quantities. It is given by
\begin{equation}
c^2_{\chi_{\rm ad}}\equiv\frac{p^\prime_\chi}{\rho^\prime_\chi}=w_\chi-\frac{w^\prime_\chi \rho_\chi}{3\mathcal{H}(1+w_\chi)\rho_\chi-Q_\chi}.
\end{equation}
The appearance of the adiabatic sound speed as well as the term $\propto \mathcal{H}^2$ is a result of a gauge transformation~\cite{Hu:1998kj,Hu:2003hjx} applied in order to relate the speed of sound in the rest frame to that in any frame and is given by
\begin{align}
\frac{\delta p_\chi}{\delta\rho_\chi}&=c^2_{s\chi}-\frac{\rho_\chi^\prime}{\delta\rho_\chi}(c^2_{s\chi}-c^2_{\chi_{\rm ad}})\frac{v_\chi-B}{k} \nonumber \\
&=c^2_{s\chi}-\frac{1}{\delta_\chi}\left[\frac{Q_\chi}{\rho_\chi (1+w_\chi)}-3\mathcal{H}\right](c^2_{s\chi}-c^2_{\chi_{\rm ad}})\frac{\Theta_\chi}{k}.
\label{dpdrho}
\end{align}
Eqs.~(\ref{dchi}) and~(\ref{tchi}) are a system of coupled equations for $\delta_\chi$ 
and $\Theta_\chi$ and the equations exibit the contributions from the interaction potential $V_3$.
Turning off the interaction term, we recover the 
evolution equations found in the literature~\cite{Hlozek:2014lca}. The corresponding equations for the DE field $\phi$ are given in Appendix~\ref{app:B}.

\section{Averaging over fast oscillations\label{sec:average}}

It is well known in the literature that solving the KG equation for potentials of the form given by Eq.~(\ref{v1}) becomes numerically intractable when $\mathcal{H}/M_\chi\ll 1$. In other words, when the period of the oscillations of the field becomes much shorter than the Hubble time $\mathcal{H}^{-1}$, the rapid oscillations can be time-averaged over one period. The equation of state of $\chi$ oscillates rapidly between $-1$ and $+1$ and therefore the averaging reveals a field redshifting as matter with $\rho_{\chi}\propto a^{-3}$ and $w_\chi=0$. The method we will follow in this work to overcome this numerical difficulty is to first solve the KG equations for DM and DE, calculate the density and pressure and their perturbations following Eqs.~(\ref{KGc0}),~(\ref{KGp0}),~(\ref{rhop})$-$(\ref{pp}),~(\ref{drhop})$-$(\ref{dpchi}),~(\ref{KGp1}) and~(\ref{KGc1}), starting from $a_{\rm ini}\sim 10^{-14}$ to the time where rapid oscillations begin which we denote by $a_{\rm osc}$. At this point we switch to solving the fluid equations for DM, i.e., Eqs.~(\ref{dchi}) and~(\ref{tchi}) while we keep tracking the evolution of the DE field $\phi$ via the KG equation. The DM fluid equations require further attention by averaging over the rapid oscillations. This time-averaging has been discussed in the literature~\cite{Li:2013nal,Poulin:2018dzj,Turner:1983he,Cembranos:2018ulm,Cembranos:2015oya} while also including DM self-interaction. Another method which avoids switching between the KG equations and the fluid equations has been proposed by refs.~\cite{Reyes-Ibarra:2010jje,Rendall:2006cq,Alho:2014fha,Jesus:2015jfa,Urena-Lopez:2015gur,Garcia-Arroyo:2024tqq}. Furthermore, ref.~\cite{Passaglia:2022bcr} has recently proposed a more accurate description of effective fluid approximation.

The main question here is how to determine, as accurately as possible, the value of $a_{\rm osc}$. We have found that up to a very good approximation, switching to the time-averaged fluid equations can be done when $M_\chi>3H$. When the transition to the fluid approximation is made, we assign the background and perturbation values of the density and pressure calculated from the KG equations as the initial values in the fluid equations.  

Now let us briefly show how the time-averaging of the background and perturbations equations is carried out. Assuming the interaction terms are small, the background KG equation for the dark matter field reads
\begin{equation}
    \chi_0^{\prime\prime}+2\mathcal{H}\chi_0^\prime+a^2 m_\chi^2\chi_0\approx 0.
    \label{kgapp}
\end{equation}
We propose as a solution to the differential equation the ansatz
\begin{equation}
    \chi_0=\chi_+(\tau)\sin\Big(\psi(\tau)\Big)+\chi_-(\tau)\cos\Big(\psi(\tau)\Big).
\end{equation}
Inserting the ansatz into Eq.~(\ref{kgapp}) and collecting terms proportional to $m_\chi$, we get
\begin{equation}
    \chi_{\pm}(\tau)=\frac{\chi^{\pm}_0}{a^{3/2}},
\end{equation}
where $\chi^{\pm}_0$ are slowly-varying functions of the conformal time. 
Therefore, the final solution looks like
\begin{equation}
    \chi_0=a^{-3/2}\Bigg[\chi_0^+(\tau)\sin\Bigg(\int a \,m_\chi\, \text{d}\tau\Bigg)+\chi_0^-(\tau)\cos\Bigg(\int a \, m_\chi \,\text{d}\tau\Bigg)\Bigg].
    \label{bkg-ansatz}
\end{equation}
Using Eq.~(\ref{bkg-ansatz}) in Eq.~(\ref{rhoc}) while still ignoring the interaction term $V_3$, the time average yields
\begin{equation}
    \langle\rho_{\chi}\rangle\simeq m_\chi^2\langle\chi_0^2\rangle.
\end{equation}
To leading order, we can drop the averages of the total time derivatives. Hence
\begin{align}
    &\langle\partial_0(\chi_0 \chi_0^\prime)\rangle=\langle\chi_0^{\prime 2}+\chi_0\chi_0^{\prime\prime}\rangle=0,
\end{align}
which allows us to write the time-averaged energy density as
\begin{eqnarray}
    \langle\rho_\chi\rangle=\Big\langle\frac{\chi_0^{\prime 2}}{2a^2}+V_{13}\Big\rangle=\Big\langle\frac{\chi_0 V_{,\chi}}{2}+V_{13}\Big\rangle.
\end{eqnarray}
where in the presence of interactions (DM self-interaction and DM-DE interaction), $V_{13}=V_1(\chi)+V_3(\phi,\chi)$. 
Now one can estimate in an efficient way, the DM equation of state using 
\begin{equation}
    w_\chi=\frac{\langle p_\chi\rangle}{\langle \rho_\chi\rangle}=\frac{\Big\langle\frac{1}{2}\chi_0 V_{,\chi}-V_{13}\Big\rangle}{\Big\langle\frac{1}{2}\chi_0 V_{,\chi}+V_{13}\Big\rangle}.
\end{equation}
A simple calculation then gives
\begin{equation}
    w_\chi=\frac{\frac{\lambda}{4}\langle\chi_0^4\rangle}{\langle\rho_\chi^0\rangle+\frac{3\lambda}{4}\langle\chi_0^4\rangle+\tilde{\lambda}\phi_0^2\langle\chi_0^2\rangle}.
\end{equation}
Using the approximation
\begin{equation}
\langle\chi_0^4\rangle\simeq\frac{3}{2}\langle\chi_0^2\rangle\langle\chi_0^2\rangle,    
\end{equation}
we finally get
\begin{equation}
    w_\chi=\frac{\dfrac{3\lambda}{8m_\chi^4}\langle\rho_\chi\rangle}{1+\dfrac{9\lambda}{8m_\chi^4}\langle\rho_\chi\rangle+\dfrac{\tilde{\lambda}\phi_0^2}{m_\chi^2}},
    \label{wchi}
\end{equation}
where $\langle\rho_\chi\rangle$ is obtained from the solution of the DM continuity equation 
$\rho_\chi^{\prime}+3\mathcal{H}(1+w_\chi)\rho_\chi=Q_\chi$. Note that in the absence of interactions, $\lambda=\tilde\lambda=0$, the equation of state is zero, representing a pressure-less fluid. However, in the presence of self-interactions the equation of state is no longer zero and interestingly, for a certain range of $\lambda$ values, $w_\chi\to 1/3$, indicating a period where the field $\chi$ behaves as radiation. We will verify this in the numerical analysis. The fact that DM self-interaction can modify the equation of state allows us to set constraints on $\lambda$. Note that the effect of the DM-DE interaction on $w_\chi$ is minimal since $\tilde\lambda$ shows up in the denominator on the right hand side of Eq.~(\ref{wchi}) for
$w_\chi$. We will see in the numerical analysis that the impact of $\tilde\lambda$ is more apparent on $w_\phi$ than it is on $w_\chi$. 

To be able to use the effective fluid equations, we still have to find the speed of sound in the DM fluid, $c_{s\chi}^2=\langle\delta p_\chi\rangle/\langle\delta \rho_\chi\rangle$. We start with the ansatz for the DM field perturbation
\begin{equation}
    \chi_1=\chi_{1}^+(k,\tau)\sin\Bigg(\int a \,m_\chi\, \text{d}\tau\Bigg)+\chi_{1}^-(k,\tau)\cos\Bigg(\int a \, m_\chi \,\text{d}\tau\Bigg),
\end{equation}
with $\chi_1^{\pm}(k,\tau)$ being slowly-varying functions of time. The averages of the density and pressure perturbations of the DM field as well as the fluid velocity are
\begin{align}
\label{dr-av}
\langle\delta\rho_\chi\rangle&=\frac{1}{a^2}\langle\chi_0^\prime\chi_1^\prime\rangle-\frac{1}{a^2}\langle\chi_0^{\prime 2}A\rangle+\langle(\Bar{V}_1+\Bar{V}_3)_{,\chi}\chi_1\rangle+\langle\Bar{V}_{3,\phi}\phi_1\rangle, \\
\label{dp-av}
\langle\delta p_\chi\rangle&=\frac{1}{a^2}\langle\chi_0^\prime\chi_1^\prime\rangle-\frac{1}{a^2}\langle\chi_0^{\prime 2}A\rangle-\langle(\Bar{V}_1+\Bar{V}_3)_{,\chi}\chi_1\rangle-\langle\Bar{V}_{3,\phi}\phi_1\rangle, \\
\frac{a^2}{k}\langle(&\rho_\chi+p_\chi)(v_\chi-B)\rangle=\langle\chi_0^\prime\chi_1\rangle.
\end{align}
Therefore we need to average every term in Eqs.~(\ref{dr-av}) and~(\ref{dp-av}). To do so we consider, to leading order, zero averaged total time derivative, i.e.,
\begin{equation}
    \Big\langle\frac{\rm d}{\text{d}\tau}\Big(\chi_0^\prime\chi_1+\chi_0\chi^\prime_1)\Big)\Big\rangle=0,
\end{equation}
which immediately gives us
\begin{equation}
    \langle\chi_0^\prime\chi_1^\prime\rangle=-\frac{1}{2}\langle\chi_0^{\prime\prime}\chi_1+\chi_0\chi_1^{\prime\prime}\rangle.
\end{equation}
After a lengthy calculation we obtain the speed of sound in the DM fluid as
\begin{equation}
    c^2_{s\chi}=\frac{\left(\dfrac{k}{2m_\chi a}\right)^2+\dfrac{3\lambda}{4m_\chi^4}\langle\rho_\chi\rangle}{1+\left(\dfrac{k}{2m_\chi a}\right)^2+\dfrac{9\lambda}{4m_\chi^4}\langle\rho_\chi\rangle+\dfrac{\tilde{\lambda}\phi_0^2}{m_\chi^2}}.
    \label{sound}
\end{equation} 
Taking the special case $\tilde{\lambda}=0$ we recover the expression of $c_{s\chi}^2$ in ref.~\cite{Cembranos:2018ulm}. Also setting $\lambda=\tilde{\lambda}=0$ reduces to the result in refs.~\cite{Park:2012ru,Hlozek:2014lca,Cembranos:2015oya,Poulin:2018dzj}.

Now that we have our complete set of equations in the general gauge, we can write them in the synchronous and conformal gauges based on Eqs.~(\ref{sync}) and~(\ref{conf}). Note that for a scalar field the anisotropic stress $\Pi$, which is the trace of $\Pi^i_j$, is zero. So from the Einstein equation
\begin{equation}
    k^2(\Psi+\Phi)=-8\pi G a^2 p \Pi,
\end{equation}
we immediately get $\Psi=-\Phi$. The equations in both gauges are summarized in Appendix~\ref{app:B}.

\section{Numerical analysis \label{sec:na}}

In this section we present the numerical results of our analysis in two ways. First we show the effect of the DM mass, DM self-interaction and DM-DE coupling on the background and perturbation observables based on a number of benchmarks. In the second way, we extract the cosmological parameters of our model using a Bayesian inference tool based on a Markov Chain Monte Carlo (MCMC) simulation. In the MCMC analysis we use different sets of data to constrain our cosmological parameters, in particular the DM mass $m_\chi$ and the couplings $\lambda$ and $\tilde\lambda$ as well as the standard parameters of $\Lambda$CDM. 

In order to evolve the background fields of DM and DE and the corresponding perturbations along with those of baryons, radiation and neutrinos, we use the Boltzmann solver \code{CLASS}~\cite{Blas:2011rf} (Cosmic Linear Anisotropy Solving System)\footnote{\href{https://github.com/lesgourg/class_public}{https://github.com/lesgourg/class\_public}} which also evolves the Einstein equations. \code{CLASS} can be interfaced with another code called \code{MontePython}\footnote{\href{https://github.com/brinckmann/montepython_public}{https://github.com/brinckmann/montepython\_public}}~\cite{Brinckmann:2018cvx,Audren:2012wb} which is a Bayesian inference tool used to sample the parameter space of our model by running a Markov Chain Monte Carlo based on the Metropolis-Hastings algorithm.

\subsection{Implementation in \code{CLASS}}

We modify the code \code{CLASS} to implement our model of interacting ultralight DM and DE fields. The modifications, which are done at the level of the background and perturbations modules, allow the user to switch from solving the KG equations to the fluid equations once the numerical evolution becomes intractable due to the rapid oscillations of $\chi$. The input parameters of the DE field include: $\mu$, the coefficient of the quintessence potential in Eq.~(\ref{v2}), the axion decay constant $F$, the initial field displacement $\phi_{\rm ini}$ and its time derivative $\phi^\prime_{\rm ini}$.  As for the DM field, the input parameters include: the DM mass $m_\chi$, the DM self-interaction $\lambda$, the DM field initial displacement $\chi_{\rm ini}$ and its time derivative $\chi^\prime_{\rm ini}$. The last free parameter is the DM-DE interaction strength $\tilde\lambda$. The integration of the equations start from $a_{\rm ini}\sim 10^{-14}$, deep in the radiation domination era. 

The initial values assigned to these input parameters are given in terms of the units adopted by \code{CLASS}.  In \code{CLASS}, fields have units of reduced Planck constant $m_{\rm Pl}=(8\pi G)^{-1/2}$ and the potential is in units of $m_{\rm Pl}^2/$Mpc$^2$ whereas the energy density is in units of Mpc$^{-2}$ and the Hubble parameter $H$ is in units of Mpc$^{-1}$. We summarize in table~\ref{tab1} the \code{CLASS} units of our parameters and their equivalent values in natural units.  

\begin{table}[H]
\centering
{\tabulinesep=1.2mm
\begin{tabu}{c|c|c}
\hline\hline
Parameter & \code{CLASS} unit & Natural unit \\
\hline
$m_\chi$ & Mpc$^{-1}$ & $6.4\times 10^{-30}$ eV  \\
\hline 
$\lambda$, $\tilde{\lambda}$ & $m_{\rm Pl}^{-2}$Mpc$^{-2}$ & $4\times 10^{-113}$ \\
\hline
$\mu$ & $m_{\rm Pl}^{1/2}$ Mpc$^{-1/2}$ & $8\times 10^{-2}$ eV  \\
\hline
$F$ & $m_{\rm Pl}$ & $10^{27}$ eV  \\
\hline\hline
\end{tabu}}
\caption{Model parameters in \code{CLASS} and natural units.}
\label{tab1}
\end{table}

For the DE field, we choose $\phi_{\rm ini}=0.05$, $\phi^\prime_{\rm ini}=1.0$ and $F=1$ (all in \code{CLASS} units). This choice indicates that initially, the kinetic term of the field dominates the potential term which means $w_\phi$ starts with a stiff matter phase, $w_\phi=1$, before transitioning quickly to $w_\phi=-1$. We have checked that changing the initial input to start with $w_\phi=-1$ instead does not impact our results. In fact, as we will see in the results section, the transition from a stiff matter phase to the DE phase happens almost instantly (with some exceptions where we saw a delayed transition but still taking place much before radiation-matter equality). For DM $\chi$, the field is initially frozen due to Hubble friction and then slowly rolls down the potential, so $\chi^\prime_{\rm ini}\sim 0$ and $\chi_{\rm ini}$ is small (in \code{CLASS} units). In fact, we don't put the initial value of $\chi^\prime_{\rm ini}$ to be strictly zero, but rather use the slow-roll approximation
\begin{equation}
    \phi^\prime_{\rm ini}\simeq \frac{a^3_{\rm ini}V(\chi_{\rm ini})}{3H_0}\,.
\end{equation}
The values of $m_\chi$, $\lambda$ and $\tilde\lambda$ are then chosen freely to explore the parameter space of the model. Now we still have to discuss the initial values taken by the parameters $\mu$ and $\chi_{\rm ini}$. The code \code{CLASS} is given the values of some parameters today, such as the DM and baryon density and the Hubble parameter $H_0$, which it tries to match by evolving the many equations governing the different species in the universe. In doing so, \code{CLASS} adjusts some parameters (that the user can provide)  by applying the ``shooting" method. We use $\mu$ and $\chi_{\rm ini}$ as our shooting parameters which are adjusted during the evolution in order to satisfy the closure relation today
\begin{equation}
    \Omega_{0\chi}+\Omega_{0\phi}+\Omega_{0b}+\Omega_{0r}=1.
\end{equation}
Here $\Omega_{0i}=\rho_i/\rho_{\rm 0,crit}, i=b,r$  and $\Omega_{0\chi}= \rho_{\chi}(1-\delta)/\rho_{\rm 0,crit}$ and $\Omega_{0\phi}= \rho_{\phi}(1-\delta)/\rho_{\rm 0,crit}$, with
$\delta=V_3/(\rho_\chi+ \rho_\phi)$. Note that the factor $(1-\delta)$ is included 
to correct for  double counting of the interaction term $V_3$. The quantity $\rho_{\rm 0,crit}$ is the critical density today which is given by
\begin{equation}
    \rho_{\rm 0,crit}=\frac{3H_0^2}{8\pi G}\,.
\end{equation}
We found that good starting values for these shooting parameters are $\mu=0.05$ and $\chi_{\rm ini}=0.1$, in \code{CLASS} units. 

As for the perturbations, we take as initial conditions, $\chi_1=\chi^\prime_1=0$ and $\phi_1=\phi^\prime_1=0$. This is analogous to setting $\delta_{\rm ini}=0$ and $\Theta_{\rm ini}=0$. Despite being set to zero initially, these quantities are quickly driven to the attractor solution~\cite{Ballesteros:2010ks}. Finally, we comment on the choice of gauge for the numerical analysis in \code{CLASS}. As mentioned in section~\ref{sec:perturb}, the synchronous gauge does not completely fix the gauge degrees of freedom and in $\Lambda$CDM we rely on the fact that $w_{\rm CDM}=0$ to fix the gauge. For a scalar field, this is not the case throughout its evolution as $w_\chi$ is dynamical. In order to be able to still consider the synchronous gauge in \code{CLASS}, we allow for a small amount of CDM by setting $\Omega_{\rm CDM} h^2=10^{-8}$. As a matter of fact, \code{CLASS} does not work if one sets $\Omega_{\rm CDM} h^2$ to zero.    

We present in the next section the results using select benchmarks to show the effect of changing the DM mass, the DM self-interaction strength and the DM-DE coupling on the background and perturbation quantities.

\subsection{Results}
\subsubsection{The effect of the dark matter mass}

\begin{figure}[H]
\begin{centering}
\includegraphics[width=0.95\textwidth]{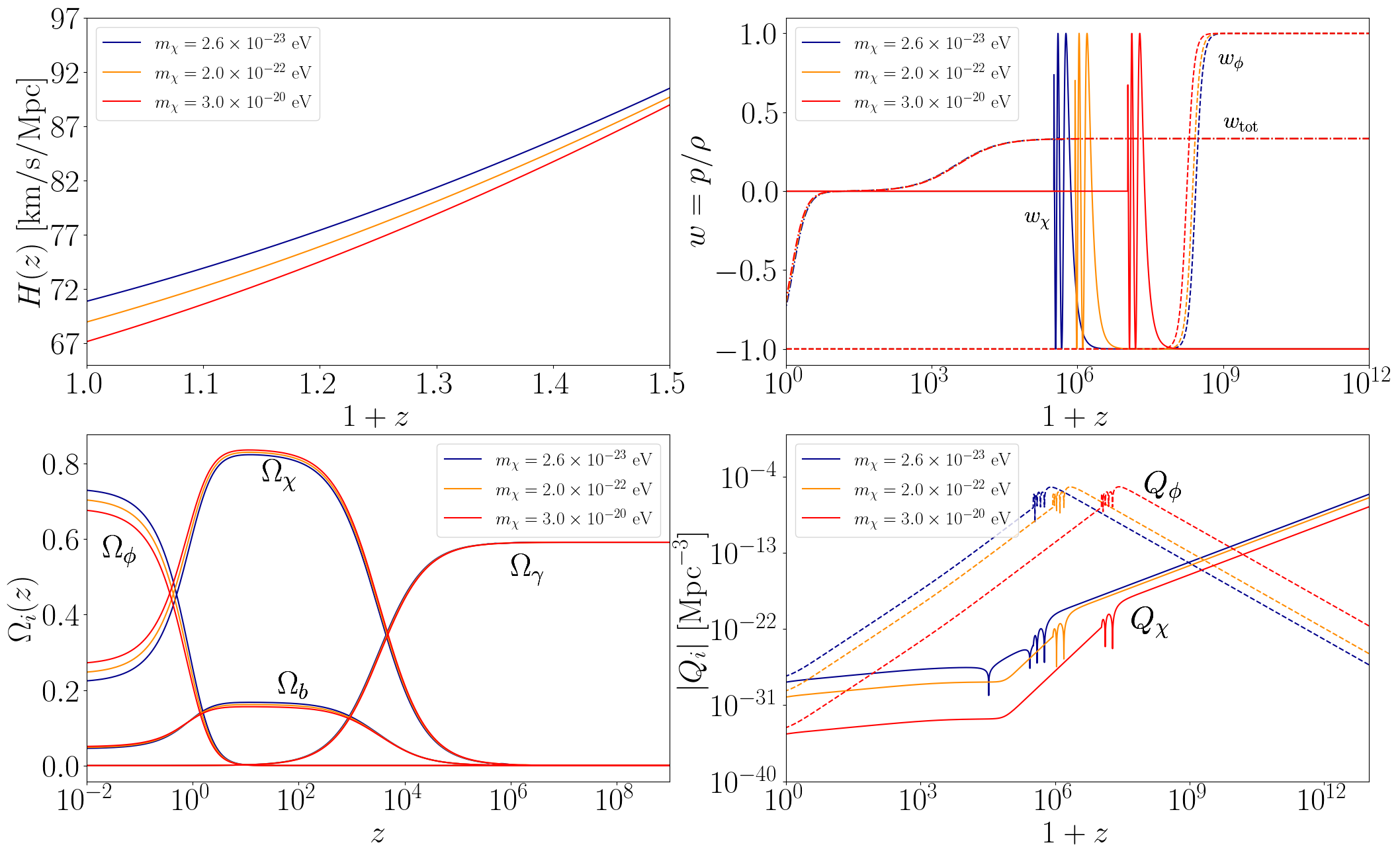}
\caption{Upper row: plot of the Hubble parameter $H(z)$ (left panel) and the DM EoS (solid), DE EoS (dashed) and the total EoS (dashdot) (right panel) versus $1+z$ for three benchmarks of the DM mass. Lower row: plots of the energy density fraction of DM, DE, baryons and radiation (left panel) and the couplings $Q_\chi$ and $Q_\phi$ (right panel) as a function of the redshift for three benchmarks of the DM mass.}
\label{fig1}
\end{centering}
\end{figure}

We begin by showing the effect of the DM mass on some background quantities. The upper left panel of Fig.~\ref{fig1} shows the evolution of the Hubble parameter $H$ as a function of the redshift for three benchmarks of DM mass. The lower left panel shows the variation of the density fractions with redshift for the same benchmarks. We fix the quantity $\theta_s$, which represents the ratio of the sound horizon to the angular diameter distance at decoupling, to its value measured accurately by the Planck experiment (i.e., $100\theta_s=1.0411)$ and ask \code{CLASS} to find the value of $\Omega_\phi$ today that will satisfy the closure relation by using the shooting method. For a light DM mass (blue curve), the model parameters can yield a universe with lower DM relic density compared to the case of a heavier DM mass (red curve). Since $\Omega_b$ and $\Omega_\gamma$ are nearly unaffected by changing the DM mass, then a lower $\Omega_\chi$ value means a higher $\Omega_\phi$ (so that the closure relation is satisfied). A universe with a higher fraction of DE allows for a larger Hubble constant $H_0$ today (blue curve).         

The upper right panel of Fig.~\ref{fig1} shows the evolution of the equation of state (EoS) for DM (solid) and DE (dashed) as well as the total EoS (dashdot). The field $\chi$ starts with $w_{\chi}=-1$ in the early universe, representing a phase of early dark energy (EDE) before undergoing rapid oscillations about $w_\chi=0$. The field then dilutes as a cold pressure-less matter similar to CDM. Heavier DM masses start to oscillate much earlier than lighter ones which makes it harder to distinguish from CDM. All the benchmarks considered are already behaving as CDM by the time of matter-radiation equality. The total EoS behaves as expected with a radiation-domination era where $w_{\rm tot}=1/3$ followed by matter domination with $w_{\rm tot}=0$ and finally a DE dominated phase with $w_{\rm tot}<-1/3$.

The bottom right panel of Fig.~\ref{fig1} shows the evolution of the couplings $Q_\chi$ and $Q_\phi$. Here the values that these couplings take are subdominant as we are considering no interactions in this section. Note the oscillatory feature visible in $Q_i$ and the turnaround during the rapid oscillations phase. We will revisit the evolution of $Q_i$ in the next section when their impacts become important.  

\begin{figure}[H]
\begin{centering}
\includegraphics[width=0.95\textwidth]{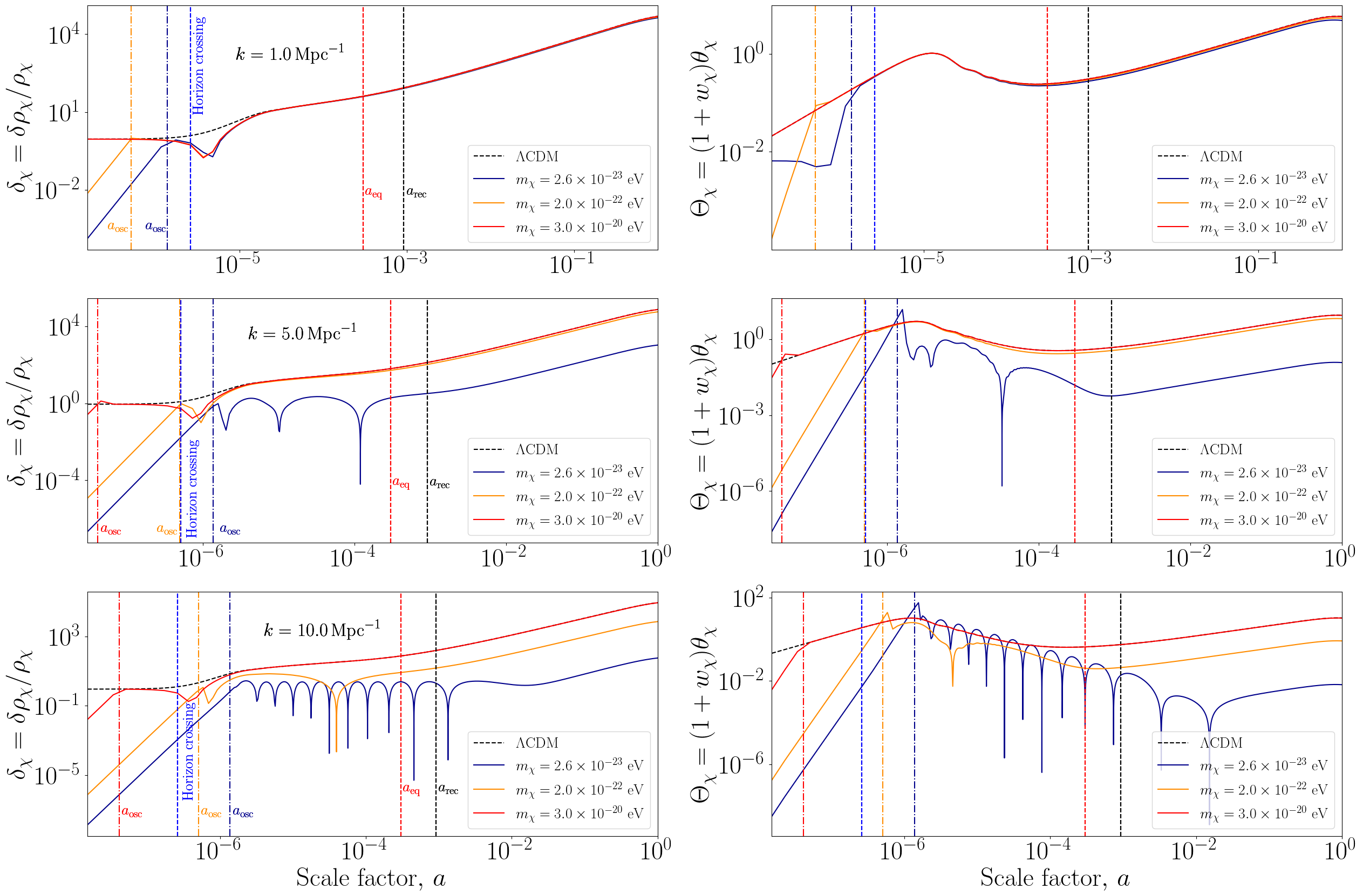}
\caption{Plots showing the DM density contrast (left) and the velocity divergence (right) as a function of the scale factor for three wavenumbers $k$. The three dotted vertical lines correspond to the time of horizon crossing (blue), matter-radiation equality (red) and recombination (black). The three dashdot vertical lines correspond to $a_{\rm osc}$, the scale factor when oscillations of the field start, with colors corresponding to each benchmark. }
\label{fig2}
\end{centering}
\end{figure}

Next, let us examine the evolution of the perturbations in the DM and DE components. We will focus mainly on the evolution of the DM density contrast $\delta_\chi$ which gives us information on structure formation. The way perturbations evolve depends on their time of horizon entry (at scale factor denoted by $a_H$), the time when the field starts its rapid oscillations and when the speed of sound starts tracking the equation of state. The growth of perturbations is controlled by an interplay between pressure and density perturbations during gravitational collapse. Therefore, the speed of sound of Eq.~(\ref{sound}) is an important quantity and leaves an imprint on the matter power spectrum. When the DM field is slowly-rolling during the EDE phase, $c^2_{s\chi}=1$, and the pressure support in the field yields a suppression of density perturbations. It is only when $c_{s\chi}^2\to w_\chi$ that the fluid is released from the pressure and the perturbations start to grow. One can understand this mechanism through a particular wavenumber defined when density and pressure perturbations are in equilibrium. This quantity is the Jeans wavenumber, $k_J$, given by
\begin{equation}
    c_{s\chi}^2k^2_J=\mathcal{H}^2\,.
\end{equation}
In the top left panel of Fig.~\ref{fig2}, the mode, $k=1.0$ Mpc$^{-1}$, is already in the CDM-like state as it enters the horizon, since $a_{\rm osc}<a_{H}$, i.e., the field has started oscillations very early on at superhorizon scale. This is true for the three benchmarks of DM masses. As the mode enters the horizon, it very quickly drops below the Jeans scale $k_J$ after a brief suppression. Once $k<k_J$, the perturbations grow, tracking exactly CDM which is shown as a black dashed line. All of this happens before matter-radiation equality. Therefore, this mode is almost indistinguishable from CDM.  The mode in the panel below it is $k=5.0$ Mpc$^{-1}$ and at least one of the three benchmarks shows a different behavior. The heavier masses (orange and red) enter the horizon already in the CDM-like phase, so they behave exactly as the $k=1.0$ Mpc$^{-1}$ case. However, the lighter mass (blue) enters the horizon while still in the EDE phase. Once the field starts oscillating for $a\geq a_{\rm osc}$ (dashdot vertical line), the perturbations also exhibit an oscillatory behavior with a constant amplitude while $k>k_J$. This suppression lasts for a while and just around the time of matter-radiation equality, $a_{\rm eq}$, the mode becomes sub-Jeans ($k<k_J$), and perturbations start to grow in the same trend as CDM but now with a clearly visible suppression compared to $\Lambda$CDM. Lastly, for the $k=10.0$ Mpc$^{-1}$ case, the three benchmarks deviate further from each other. The heavier mass (red), just like the above two cases, is still CDM-like by the time it crosses the horizon. The mode becomes sub-Jeans almost immediately and tracks the CDM growth for $a\gtrsim 10^{-6}$. The other two masses (blue and orange) enter the horizon in the EDE phase, i.e., with $w_\chi=-1$. In this case, the pressure support in the field is still strong and the perturbations become suppressed compared to their evolution at superhorizon scale. In other words, for $k>k_J$ the two benchmarks experience suppression of growth and as $w_\chi\to 0$, the field starts to oscillate with an almost constant amplitude. The mode for the intermediate mass (orange) drops below the Jeans scale before the lighter mode does. The intermediate mode then grows and trends as CDM but with a visible suppression. The lighter mode continues oscillating for a longer period of time resulting in a larger suppression of growth. Once $k<k_J$, the mode starts to grow but with further suppression in comparison with the other two benchmarks. 

The same observations can be made for the velocity divergence in the right panels of Fig.~\ref{fig1}, where the suppression of growth is visible for the scalar DM case in comparison to CDM owing to the fact that an ultralight scalar field has a characteristic Jeans scale. One effect that distinguishes the evolution of $\Theta_\chi$ from $\delta_\chi$ is the fact that we have a decaying amplitude for $\Theta_\chi$ during oscillations, opposed to a constant amplitude for $\delta_\chi$.

\begin{figure}[H]
\begin{centering}
\includegraphics[width=0.95\textwidth]{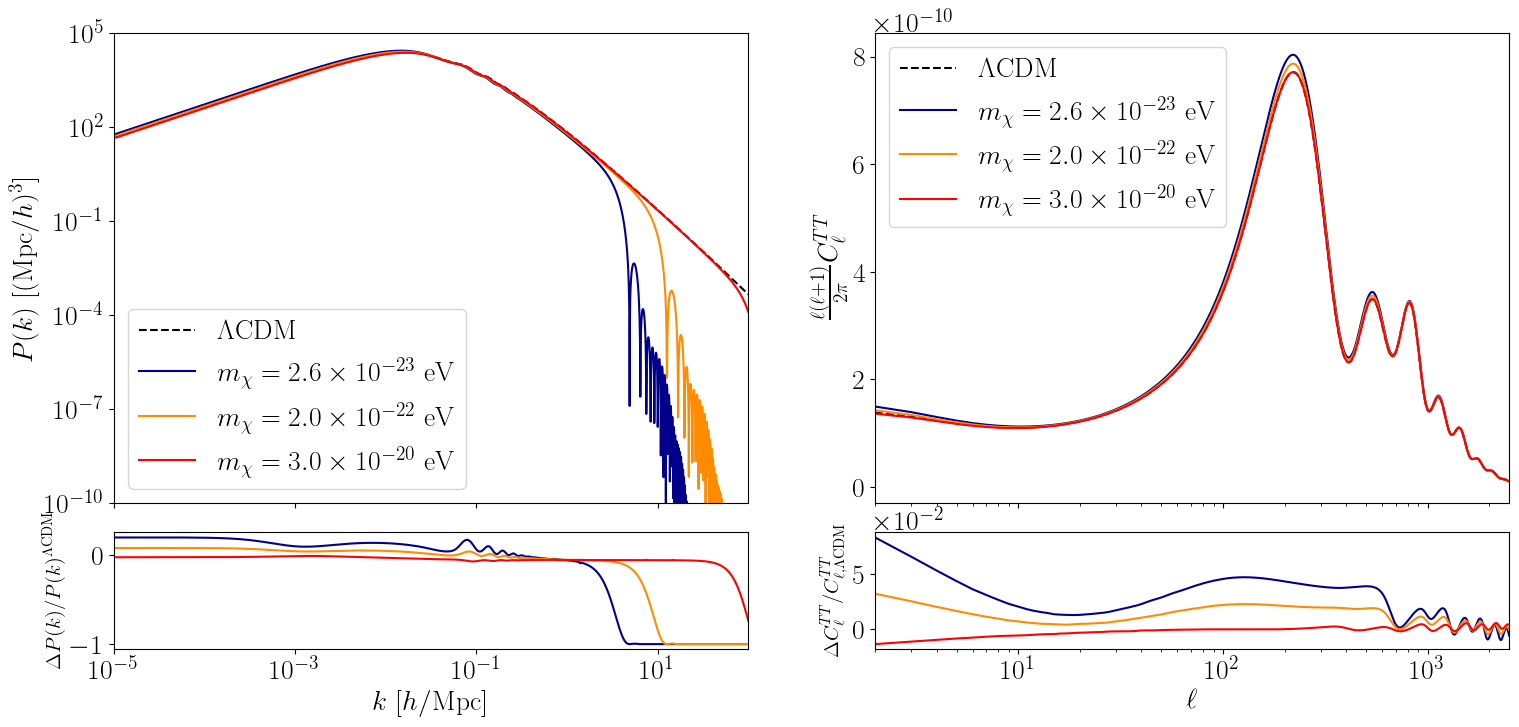}
\caption{Left panel: the matter power spectrum plotted against the wavenumber for three benchmarks of DM mass. Right panel: the temperature TT power spectrum as a function of the multipoles also for three benchmarks of DM mass. The dashed line represents $\Lambda$CDM. }
\label{fig3}
\end{centering}
\end{figure}

The evolution of the DM perturbations shown in Fig.~\ref{fig2} are reflected in the matter power spectrum displayed in the left panel of Fig.~\ref{fig3}. The spectrum has a characteristic peak at $k_{\rm eq}=a_{\rm eq}H(a_{\rm eq})$ corresponding to the mode entering the Hubble radius at the time of matter-radiation equality. The power $P(k)\propto k^{n_s}$ for $k<k_{\rm eq}$ and $\propto k^{n_s-4}$ for $k>k_{\rm eq}$, where $n_s$ is the spectral index constituting one of the free parameters of $\Lambda$CDM. For large scales ($k<k_{\rm eq}$), the ultralight scalar field DM tracks CDM with little to no deviation, while at small scales ($k>k_{\rm eq}$), the suppression of power due to the presence of a scale-dependent growth for light scalar fields is evident in the cutoff at large $k$ values in the matter power spectrum. Lighter DM shows the strongest suppression while heavier DM becomes almost indistinguishable from CDM.

In the right panel of Fig.~\ref{fig3} we show the temperature power spectrum for the three benchmarks along with $\Lambda$CDM. Changing the DM mass affects the DM density as Fig.~\ref{fig1} suggests, with the lighter mass having the smallest DM abundance. Lowering the DM content decreases the DM-to-photon and DM-to-baryon ratios (for a fixed amount of baryons) which causes the overall amplitude of the peaks to increase due to the enhancement of radiation driving, an effect clearly visible in Fig.~\ref{fig3}. Furthermore, the position of the first acoustic peak is not changed and this is because of the requirement of a fixed $\theta_s$ and a changing $H_0$. As a consequence, the (integrated) Sachs-Wolfe [(I)SW] plateau at low $\ell$ shows major deviation from $\Lambda$CDM as is evident from the smaller plot of the relative change in the temperature power spectrum, $\Delta C^{TT}_{\ell}/C^{TT}_{\ell,\Lambda\text{CDM}}$.

\subsubsection{The effect of dark matter-dark energy interaction}

\begin{figure}[H]
\begin{centering}
\includegraphics[width=0.95\textwidth]{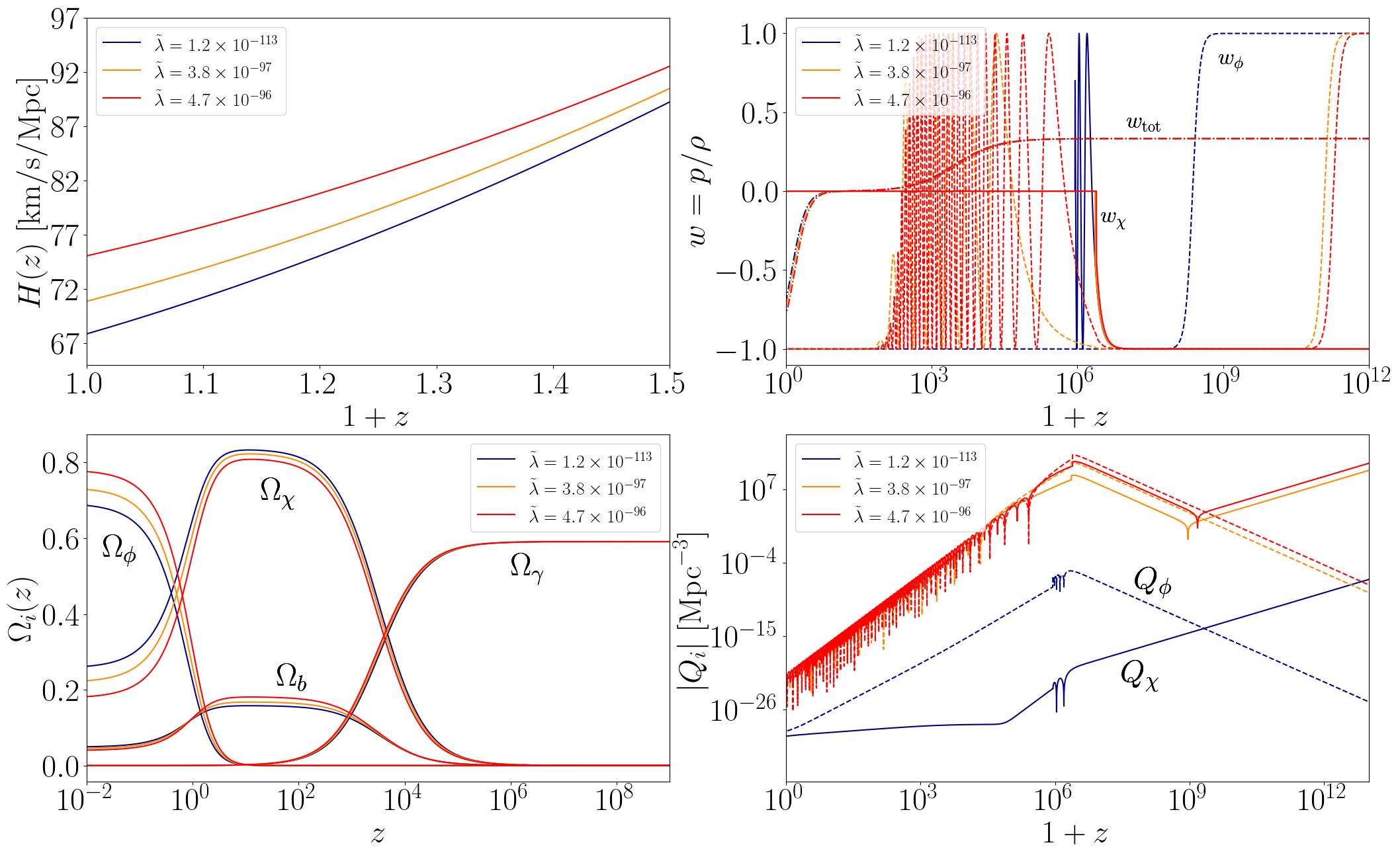}
\caption{Same as in Fig.~\ref{fig1} but for benchmarks representing three DM-DE interaction strengths.  }
\label{fig4}
\end{centering}
\end{figure}

\begin{figure}[H]
\begin{centering}
\includegraphics[width=0.95\textwidth]{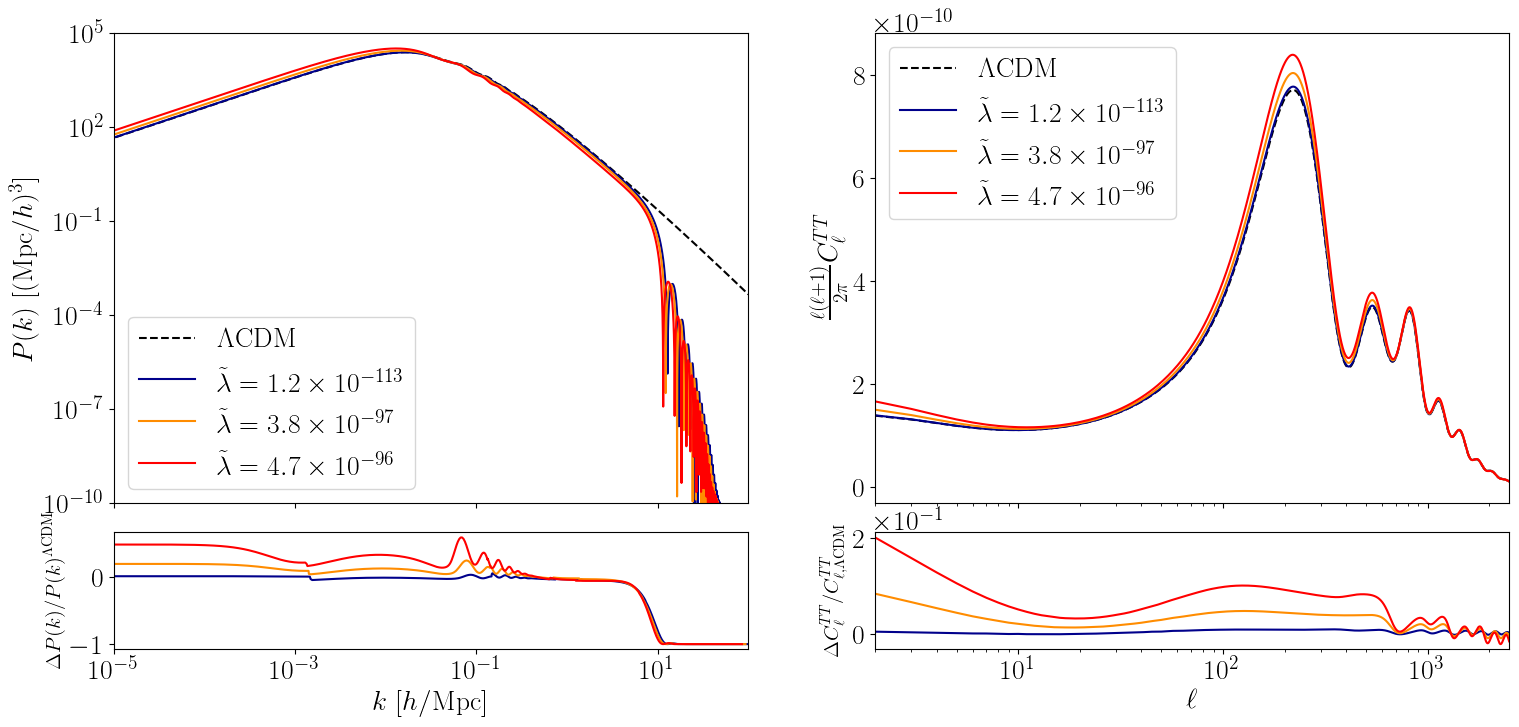}
\caption{Same as in Fig.~\ref{fig3} but for benchmarks representing three DM-DE interaction strengths.}
\label{fig5}
\end{centering}
\end{figure}

In order to elucidate the effect of DM-DE interaction on cosmology, we will fix the DM mass in this section to $m_\chi=2.0\times 10^{-22}$ eV and turn off the DM self-interaction. The parameter controlling the DM-DE interaction strength is $\tilde\lambda$ for which we choose three values as shown in the legends of Fig.~\ref{fig4}. The Hubble parameter today $H_0$ is affected by $\tilde\lambda$ and an increase in this parameter requires smaller DM density fraction $\Omega_\chi$ in order to keep $\theta_s$ fixed to its Planck value which renders a higher $H_0$ value. As mentioned before, a smaller $\Omega_\chi$ means a larger $\Omega_\phi$ which drives the accelerated expansion of the universe at a higher rate, i.e.,  a larger $H_0$ value. Again, we see that a smaller DM-to-baryon ratio for larger $\tilde\lambda$ increases the size of the acoustic peaks as visible in the right panel of Fig.~\ref{fig5}. Changes in the DM-to-baryon and DM-to-photon ratios affect the time of matter-radiation equality which is constrained by Baryon Acoustic Oscillations (BAO) measurements from the CMB. Therefore, $\tilde\lambda$ will be subject to this constraint as we will see in the next section.  

The DM, DE and total equations of state are plotted in the top right panel of Fig.~\ref{fig4} for the three benchmarks of DM-DE interaction. The interaction strength has little effect on $w_\chi$ where the transition from 
EDE {phase} to CDM-like phase happens at almost the same redshift for the different values of $\tilde\lambda$. The reason for this can be inferred from Eq.~(\ref{wchi}), where the $\tilde\lambda$ term only appears in the denominator and with $\lambda=0$, the EoS $w_\chi=0$, and so $\tilde\lambda$ has no effect. The same cannot be said about $w_\phi$, where the duration of the early stiff matter-like phase ($w_\phi=1$) is strongly reduced for larger interaction strengths. In this case, the drop to $w_\phi=-1$ happens almost instantly as one can clearly see. {This is because the large interaction strength causes an increase in the potential, thus overwhelming the kinetic term in the pressure and density equations, leading to $w_\phi\to -1$. } Moreover, another effect is that DM oscillations are inherited by DE at a later time where $w_\phi$ exhibits rapid oscillations with decaying amplitude before reaching $w_\phi=-1$. \\
Now let us examine the evolution of the couplings $Q_\chi$ and $Q_\phi$ shown in the lower right panel of Fig.~\ref{fig4} (in their absolute values). Once the interaction strength $\tilde\lambda$ is switched on, the couplings $Q_\chi$ (solid curve) and $Q_\phi$ (dashed curve) significantly increase by more than 11 orders of magnitude. The coupling $Q_\phi$ increases steadily with $z$ until $z\sim 10^6$ where it turns around and decreases sharply. This is so because
\begin{equation}
    Q_\phi=\tilde\lambda \phi_0^2\chi_0\chi^\prime_0\simeq\frac{\tilde\lambda \phi_0^2}{2m_\chi^2}\langle\rho_\chi\rangle^\prime\,,
\end{equation}
where $\langle\rho_\chi\rangle^\prime$ drops as $a^{-4}$. The coupling $Q_\chi$ on the other hand, starts to decrease at early times and the small spike seen around $z\sim 10^9$ is due to the fact that $Q_\chi$ has turned negative at this point. At $z\sim 10^9$, $Q_\chi$ coupling starts to drop but not as fast as $Q_\phi$ since
\begin{equation}
    Q_\chi=\tilde\lambda \phi_0\phi^\prime_0\langle\chi_0^2\rangle \simeq\frac{\tilde\lambda \phi_0\phi^\prime_0}{m_\chi^2}\langle\rho_\chi\rangle\,,
\end{equation}
where $\langle\rho_\chi\rangle\sim a^{-3}$. Both couplings remain sizable till recombination and have the same sign for most of their evolution. The presence of a source term $Q_\chi$ in the continuity equation of $\rho_\chi$ adds an extra contribution that evolves as $a^{-3}$ resulting in an overall decrease in $\rho_\chi$. This means that $\Omega_\chi$ today must be smaller. The heights of the CMB acoustic peaks fix the DM to baryon ratio and the only way to change this ratio by making it smaller without violating the Planck measurements is to increase $H_0$. This shows us how a DM-DE interaction can lead to an enhancement in $H_0$. Similar to the case of changing the DM mass discussed in the previous section, we have fixed $\theta_s$ and allowed \code{CLASS} to determine $H_0$. The impact that this has on the temperature power spectrum is visible in the (I)SW plateau as seen in the right panel of Fig.~\ref{fig5}. 

The effect on DM-DE interaction on the matter power spectrum is not significant. In the left panel of Fig.~\ref{fig5}, the three curves corresponding to the three $\tilde\lambda$ benchmarks nearly coincide at the cutoff tail but all deviate significantly from $\Lambda$CDM. In fact this deviation from $\Lambda$CDM is mainly due to the DM mass. However, we see slight deviation between the three benchmarks at large scales (for $k<k_{\rm eq}$) for reasons that will become clear when discussing the evolution of perturbations. 

\begin{figure}[H]
\begin{centering}
\includegraphics[width=0.95\textwidth]{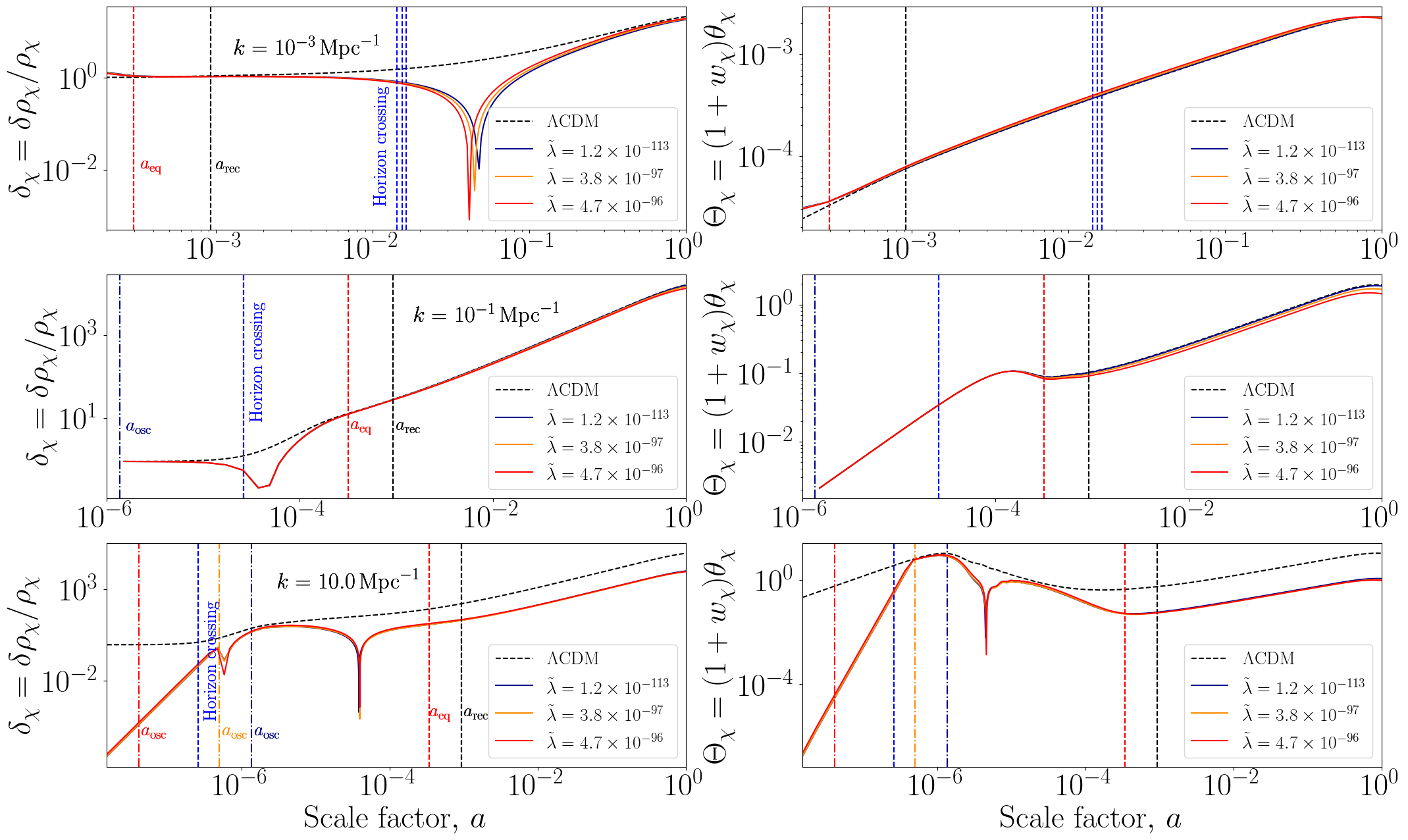} 
\caption{Same as in Fig.~\ref{fig2} but for benchmarks representing three DM-DE interaction strengths.}
\label{fig6}
\end{centering}
\end{figure}

In Fig.~\ref{fig6}, we plot the evolution of the density contrast $\delta_\chi$ and the flux (velocity divergence) $\Theta_\chi$ for the three DM-DE interaction benchmarks. The top and middle panels correspond to the modes $k=10^{-3}$ Mpc$^{-1}$ and $k=10^{-1}$ Mpc$^{-1}$, respectively. One can see that the density contrast tracks $\Lambda$CDM at superhorizon scale and only begins to depart close to horizon entry. The scale-dependent growth due to the Jeans scale takes over at subhorizon scale, where suppression of growth is clearly visible at the beginning. But almost immediately after $k<k_J$, the perturbations grow, tracking $\Lambda$CDM again. In the top panel, DM-DE interaction has minimal effect on the mode at superhorizon scale (the three curves overlap) but the effect becomes visible after horizon entry where the perturbations corresponding to different interaction strengths separate before increasing and tracing $\Lambda$CDM again. This effect is imprinted in the matter power spectrum for $k<k_{\rm eq}$ seen in the left panel of Fig.~\ref{fig5}. Higher modes, i.e., $k=10^{-1}$ Mpc$^{-1}$ (middle panel) and $k=10$ Mpc$^{-1}$ (bottom panel), do not seem to be affected much by DM-DE interaction as the three curves overlap. However, they still deviate from $\Lambda$CDM, especially higher modes. For $k=10$ Mpc$^{-1}$, the mode starts to oscillate as it enters the horizon with $k>k_J$ causing a suppression of growth. Once the mode becomes sub-Jeans, the pressure in the fluid drops and the perturbations grow, trending in the direction of $\Lambda$CDM while remaining suppressed in comparison to CDM.   {The fact that $\tilde\lambda$ does not drastically impact perturbations comes from its minimal effect on the speed of sound. By examining Eq.~(\ref{sound}), the effect of a large $\tilde\lambda$ for the case when $\lambda=0$ depends on the size of the mode $k$. For small $k$, the impact of $\tilde\lambda$ becomes visible as opposed to the case of large $k$ where the effect of $\tilde\lambda$ is diluted. This is the reason for the slight shift between the curves in the top left panel of Fig.~\ref{fig6} and the complete overlap for larger modes. }

\subsubsection{The effect of dark matter self-interaction}

\begin{figure}[H]
\begin{centering}
\includegraphics[width=0.95\textwidth]{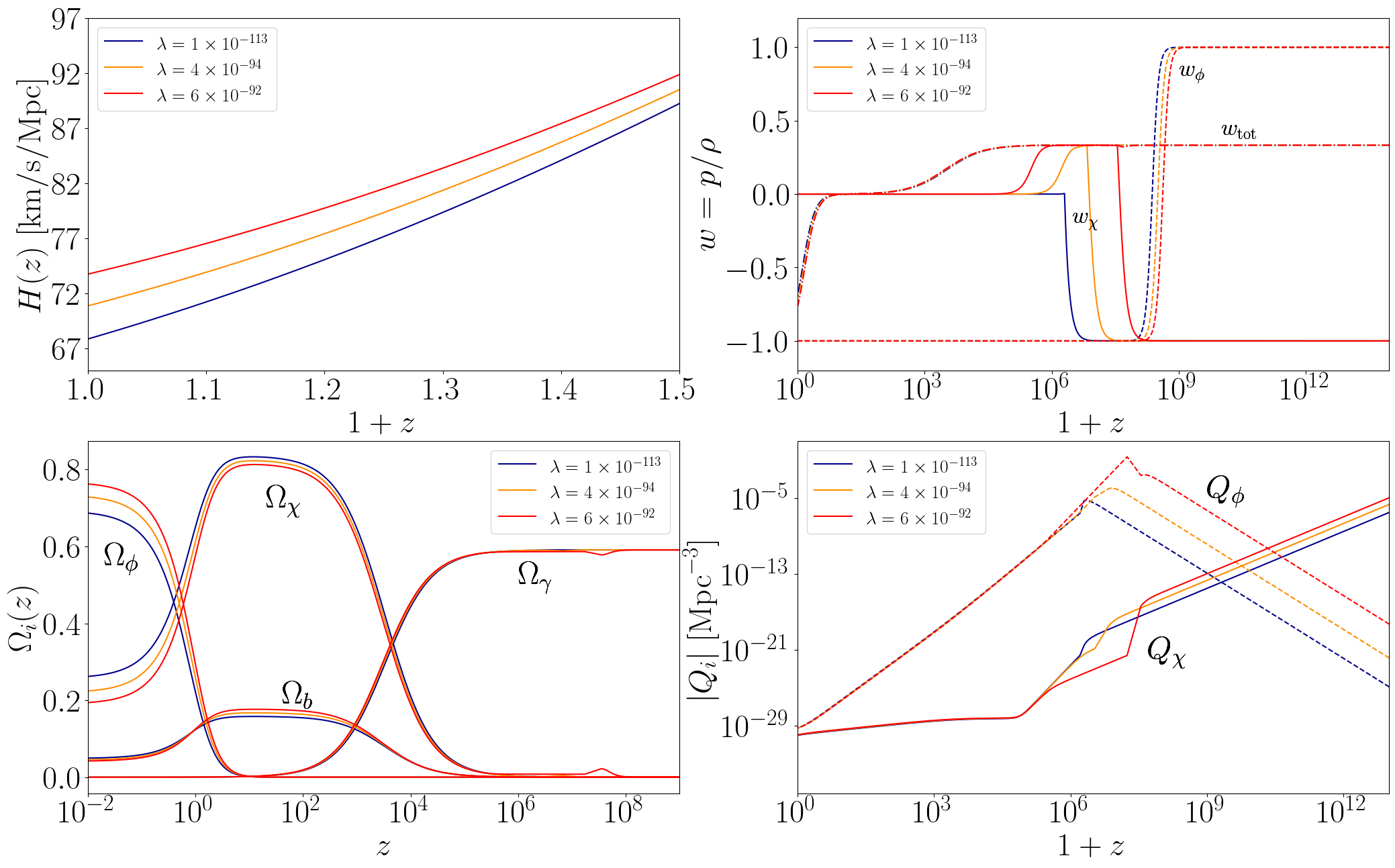}
\caption{Same as in Fig.~\ref{fig1} but for benchmarks representing three DM self-interaction strengths.}
\label{fig7}
\end{centering}
\end{figure}

\begin{figure}[H]
\begin{centering}
\includegraphics[width=0.95\textwidth]{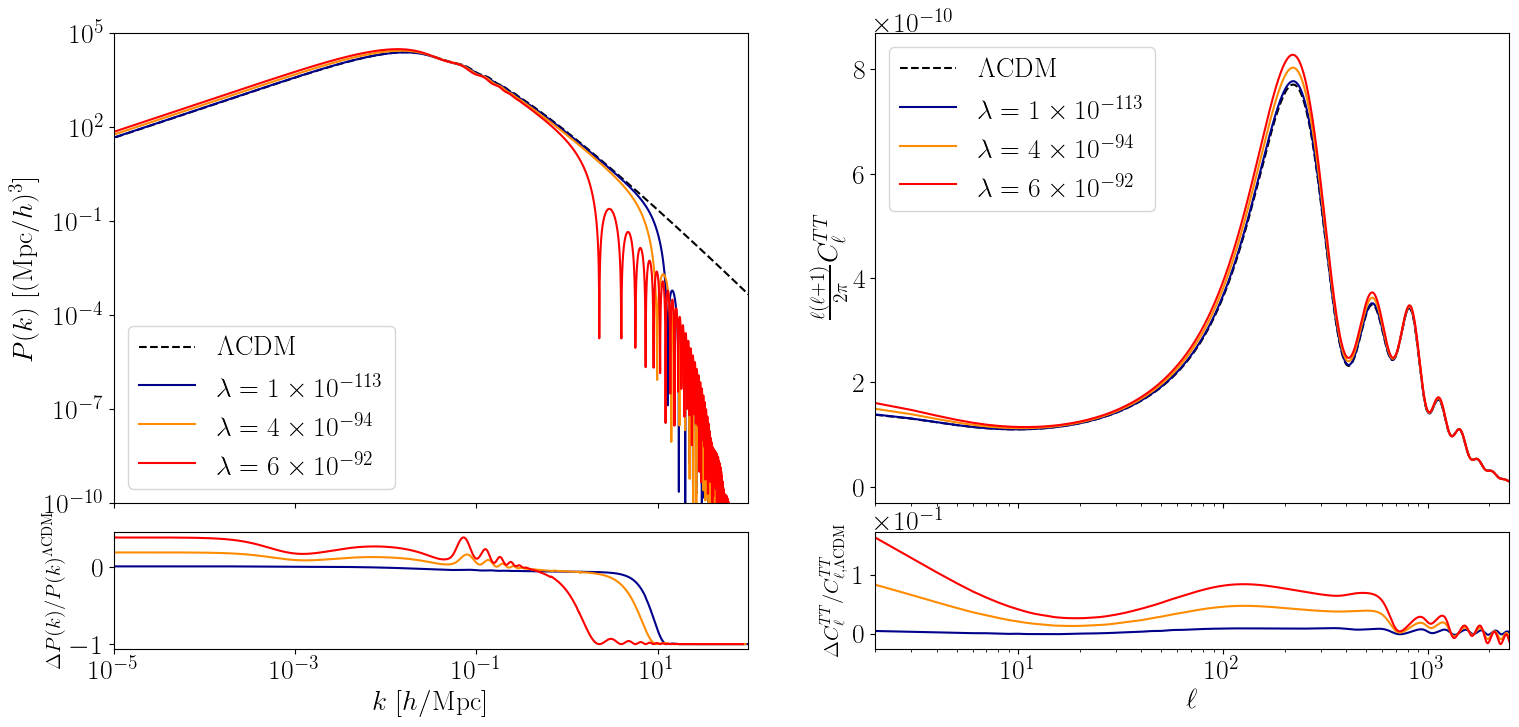}
\caption{Same as in Fig.~\ref{fig2} but for benchmarks representing three DM self-interaction strengths.}
\label{fig8}
\end{centering}
\end{figure}

In this section we will study the effect of DM self-interaction controlled by the parameter $\lambda$. Here we will fix the DM mass to $m_\chi=2.0\times 10^{-22}$ eV and switch off DM-DE interaction. We observe that DM-DE interaction and DM self-interaction have the same effect on the Hubble parameter and on the DM and DE density fractions as shown in the top left and bottom left panels of Fig.~\ref{fig7}. The extra contribution to the DM energy density coming from the self-interaction term in the potential $V_1(\chi)$ can be made consistent with the Planck measurements, i.e. keeping $\theta_s$ fixed, by lowering the value of $\Omega_\chi$ today. In this case the DM-to-baryon ratio decreases causing an enhancement in the acoustic peaks of the temperature power spectrum (see right panel of Fig.~\ref{fig8}). This is compensated by an increase in $H_0$ which is also reflected in an increase in $\Omega_\phi$ and a change in the Sachs-Wolfe plateau. DM self-interaction can also impact the couplings $Q_\chi$ and $Q_\phi$, though not directly. The modest increase in the values of $Q$ seen in the bottom right panel of Fig.~\ref{fig7} can be explained by the effect of $\lambda$ on the fields $\chi$ and $\phi$ following the solution of the KG equation, as well as $\rho_\chi$ from the continuity equation.    

An interesting effect that DM self-interaction has on background fields comes from the DM equation of state. One can see from Eq.~(\ref{wchi}) that for strong DM self-interaction, i.e. for $9\lambda\langle\rho_\chi\rangle/8m_\chi^4\gg 1$, we have $w_\chi\simeq 1/3$. This effect can be clearly seen in the upper right panel of Fig.~\ref{fig7} where the red and orange solid curves plateau at $w_\chi\simeq 1/3$ for a period of time. This means that after the EDE phase, the field $\chi$ enters a radiation-like period before the EoS falls to $w_\chi=0$, where it now behaves as CDM. As for DE, the EoS $w_\phi$ changes very slightly due to the coupled nature of the DM and DE background equations. Note that here we kept $\tilde\lambda$ at a very small value but not exactly zero.

DM self-interaction has a clear impact on the matter power spectrum as shown in the left panel of Fig.~\ref{fig8}. Unlike DM-DE interaction, DM self-interaction affect small scales (as well as large scales in a manner similar to DM-DE interaction). The reason power at small scales (large $k$, i.e., $k>k_{\rm eq}$) is affected by $\lambda$ comes from the strong dependence of the sound speed $c_{s\chi}^2$ on $\lambda$, as given by Eq.~(\ref{sound}). For large $k$, the speed of sound remains sizable thus sustaining pressure in the fluid. This leads to suppression of power as one can clearly see from the cutoff in the matter power spectrum for large $k$.  

\begin{figure}[H]
\begin{centering}
\includegraphics[width=0.95\textwidth]{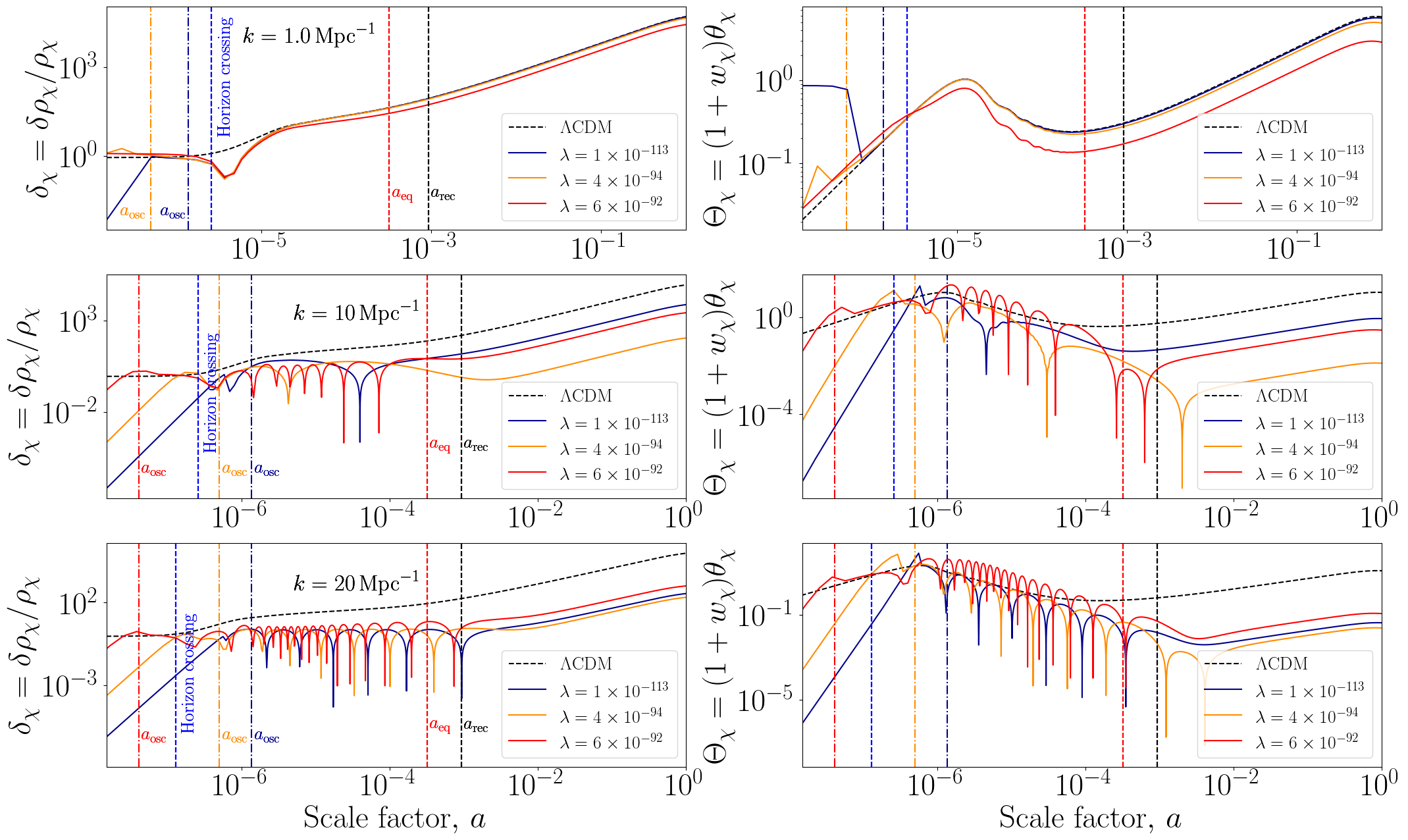}
\caption{Same as in Fig.~\ref{fig3} but for benchmarks representing three DM self-interaction strengths.}
\label{fig9}
\end{centering}
\end{figure}

Let us now examine closely the evolution of DM density perturbations which can explain the pattern seen in the matter power spectrum. For the mode $k=1.0$ Mpc$^{-1}$ in the top left panel of Fig.~\ref{fig9}, the perturbation crosses the horizon already in the CDM-like phase ($a_H>a_{\rm osc}$) and almost immediately reaches sub-Jeans scale, i.e., $k<k_J$. The perturbation then grows and starts tracing $\Lambda$CDM with the only slight deviation coming from the benchmark with the largest self-interaction strength $\lambda$. The reason can be attributed to the speed of sound where larger $\lambda$ means higher sound speed. This translates to a higher pressure in the fluid and so suppression is maintained for a longer period of time before the mode becomes sub-Jeans. This is what is also seen in the middle and bottom panels for $k=10$ Mpc$^{-1}$ and $k=20$ Mpc$^{-1}$. Naturally, higher modes have larger sound speed and with DM self-interaction switched on, an enhancement in the sound speed is obtained causing the modes to oscillate with a constant amplitude after horizon entry. Notice how the red curve oscillates at a higher frequency than the orange curve which is due to a larger $\lambda$ and therefore a larger pressure in the fluid. In the middle panel, the mode with the largest $\lambda$ has already reached a CDM-like phase prior to horizon crossing. At superhorizon scale, the perturbation is very close to $\Lambda$CDM and as it enters the horizon, it begins to oscillate at a constant amplitude while it remains super-Jeans ($k>k_J$). This mode becomes sub-Jeans before matter-radiation equality, after which oscillations cease and the perturbation starts growing and trending in the direction of $\Lambda$CDM. Still focusing on the benchmark with largest $\lambda$ (red curve), the bottom panel shows a longer period of oscillations because of a larger sound speed. The mode becomes sub-Jeans around matter-radiation equality, but oscillations continue for a while with growing amplitude. In all three panels, the red curve exhibits a suppression of perturbations in comparison to $\Lambda$CDM. For the first two benchmarks (blue and orange curves), the mode $k=1.0$ Mpc$^{-1}$ in the upper left panel behaves similarly to the benchmarks with largest $\lambda$ but with no noticeable suppression in comparison to $\Lambda$CDM. In the middle and bottom panels, the modes corresponding to the first two benchmarks enter the horizon while still in either their EDE phase or radiation-like phase. Growth becomes suppressed in comparison to superhorizon evolution and after the modes enter their CDM-like phase (when $a_{\rm osc}>a_H$), oscillations ensue causing growth suppression as long as $k>k_J$. Once the mode becomes sub-Jeans, the perturbation grows but with a noticeable suppression in comparison with $\Lambda$CDM.  

The fact that DM self-interaction impacts the sound speed more strongly than the DM-DE interaction {suggests that cosmological observations can constrain $\lambda$ and $\tilde\lambda$ to varying degrees}. In other words, constraints on the DM self-interaction strength may come from both background and perturbation observables while DM-DE interaction strength will be mostly constrained by background observables.

\subsection{Constraints from cosmological observations}

The benchmark values chosen in the previous section were for the sole purpose of showing the effect of the DM mass, DM self-interaction and DM-DE interaction on cosmology. In this section we will conduct an extensive statistical analysis of the model parameter space using cosmological data sets to try and constrain the free model parameters, $m_\chi$, $\lambda$ and $\tilde\lambda$. The data sets used in our analysis are as follows:

\begin{enumerate}
    \item The Planck 2018 temperature anisotropies and polarization measurements. The temperature and polarization (TT TE EE) likelihoods include low multipole data ($\ell<30$)~\cite{Planck:2018vyg,Planck:2018nkj,Planck:2019nip}. The high multipole likelihood includes: $30\lesssim \ell\lesssim 2500$ for the TT spectrum and $30\lesssim \ell\lesssim 2000$ for the TE and EE spectra. The low-E polarization likelihood includes $2\leq \ell\leq 30$ for the EE spectrum.  
    \item The Planck 2018 lensing likelihood~\cite{Planck:2018lbu} which is inferred from the lensing potential power spectrum.
    \item Baryon Acoustic Oscillation (BAO) data gathered by the Sloan Digital Sky Survey (SDSS) which includes the data releases: the DR7 Main Galaxy Sample~\cite{Ross:2014qpa}, the DR9 release~\cite{BOSS:2012bus}, the Baryon Oscillation Spectroscopic Survey (BOSS) DR12 survey~\cite{BOSS:2016wmc} and the SDSS improved final results spanning eight different redshift intervals~\cite{eBOSS:2020yzd}. We also include the BAO+full shape likelihood for the SDSS DR7 Main Galaxy Sample (MGS)~\cite{Howlett:2014opa} and the 6dF Galaxy Survey~\cite{Beutler:2011hx}.
    \item The combination Pantheon+SH0ES~\cite{Brout:2022vxf,Riess:2021jrx} data set which uses an additional Cepheid distance as a calibrator of the Supernova SNIa intrinsic magnitude. 
    \item For Large Scale Structure (LSS) data, we use the WiggleZ survey~\cite{Parkinson:2012vd} which measures the galaxy power spectrum in four bins of redshift centered at $z=0.22, 0.41, 0.60$ and $0.78$. We only consider scales up to $k\lesssim 0.2h$ Mpc$^{-1}$ to minimize the non-linear effects which we do not take into consideration. We should note that the BOSS analysis~\cite{BOSS:2016psr} takes into account mild non-linear effects even at $k=0.15h$ Mpc$^{-1}$. Thus our result using the WiggleZ data should be considered preliminary pending a further analysis including non-linear corrections. 
 
\end{enumerate}

\begin{figure}[H]
\begin{centering}
\includegraphics[width=0.95\textwidth]{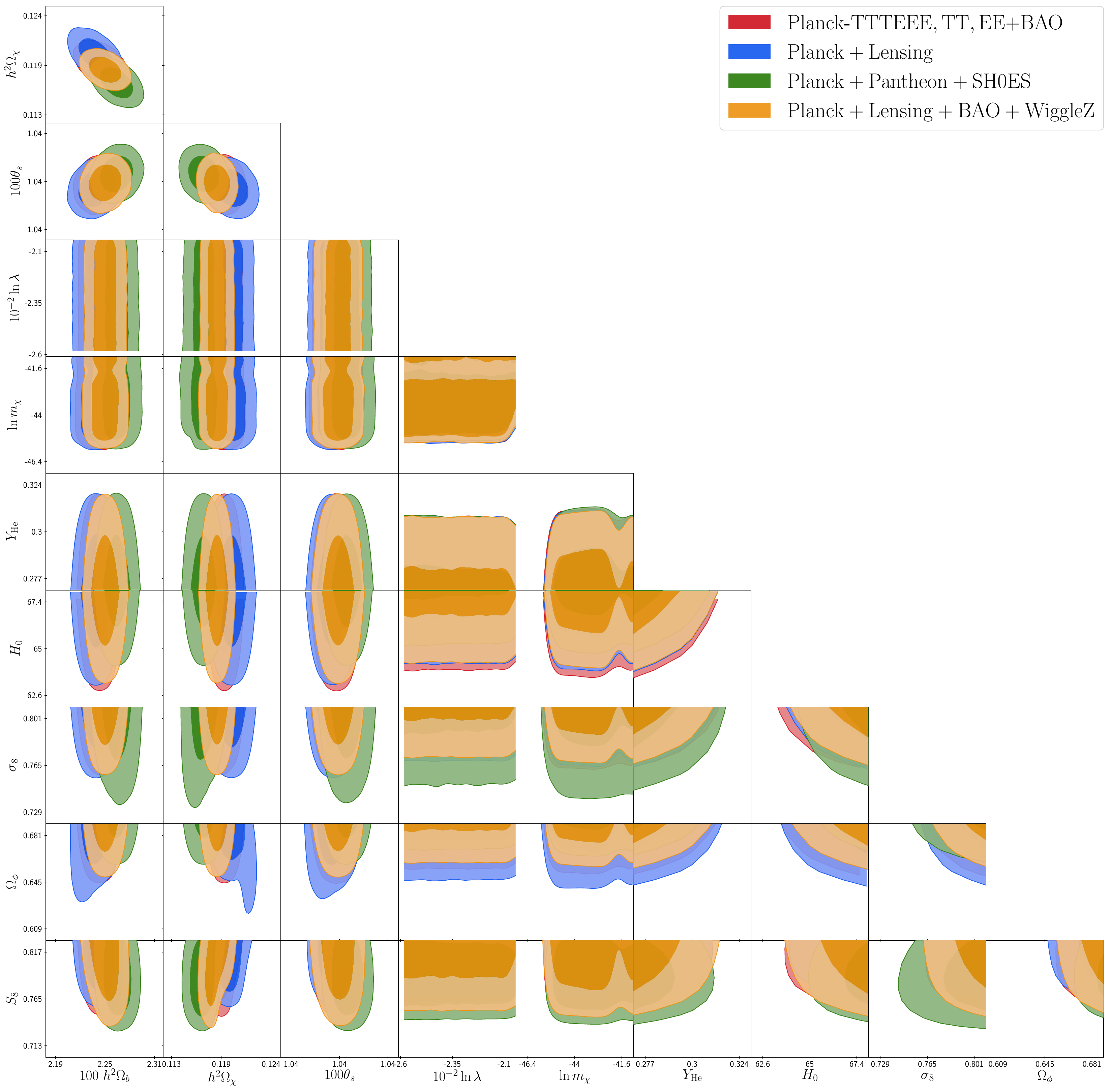} 
\caption{The triangular posterior distributions of some of our model cosmological parameters for a combination of datasets shown in the figure legend. For each dataset, we show the allowed regions at 68\% and 95\% CL. }
\label{fig10}
\end{centering}
\end{figure}

We use the MCMC sampler \code{MontePython} with the Metropolis-Hastings algorithm  to extract constraints on the cosmological parameters with the above data sets. The sampling parameters consist of the baseline $\Lambda$CDM parameters along with the three additional free parameters of our model
\begin{equation}
    \underbrace{\Omega_b h^2, ~~\Omega_\chi h^2, ~~z_{\rm reio}, ~~\theta_s, ~~A_s, ~~n_s}_{\Lambda\text{CDM}}, ~~m_\chi, ~~\lambda, ~~\tilde\lambda\,,
\end{equation}
where $\Omega_c h^2$ is replaced by $\Omega_\chi h^2$ which is our DM field, $A_s$ is the amplitude of primordial fluctuations and $z_{\rm reio}$ is the redshift at reionization. We choose flat priors for the $\Lambda$CDM parameters and logarithmic priors for our three model parameters. {We also adopt the convention of the Planck collaboration in choosing free-streaming neutrinos as two massless species and one massive with $m_\nu = 0.06$ eV~\cite{SimonsObservatory:2018koc}.}

The above parameters are not the only ones as there are  many nuisance parameters involved especially those from the Planck likelihoods. \code{MontePython} uses the Cholesky decomposition~\cite{Lewis:2013hha} of the covariance matrix which helps with convergence in the presence of a large number of nuisance parameters. We monitor the convergence of the chains using the Gelman-Rubin~\cite{Gelman:1992zz} criterion $R-1<0.05$. Our derived parameters are the Hubble parameter $H_0$, $\sigma_8$ (and $S_8$), the DE density fraction $\Omega_\phi$ and the total matter density fraction $\Omega_{\rm m}$. 

We plot in Fig.~\ref{fig10} the 2D posterior distributions of some of the sampling and derived parameters of our model for a different combinations of the considered data sets as shown in the figure legend. We can identify some strong correlations between some of the parameters. There is a positive correlation between $\Omega_\chi$ and $S_8$ and a negative correlation between $H_0$ and $S_8$ and a positive one between $H_0$ and $\Omega_\phi$. This is to be understood since higher $H_0$ values require smaller DM density and so a smaller $S_8$ and larger DE density. We only see a slight correlation between $H_0$ and $\lambda$ for larger DM self-interaction strengths. The DM mass-$\lambda$ plane shows a slight bend for larger $\lambda$ value which will be the origin of the constraint on $\lambda$. Notice that even though we sampled the lower DM mass region, the posterior ends up not favoring this mass range, considering $\chi$ as comprising the entire DM density today.    

\begin{table}[H]
\centering
{\tabulinesep=1.2mm
\resizebox{\textwidth}{!}{\begin{tabu}{cccccc}
\hline\hline
Parameter & Planck & Planck & Planck+Pantheon & Planck+Lensing & ALL \\
& +BAO & +Lensing & +SH0ES & +BAO+WiggleZ \\
\hline
100$\Omega_b h^2$ & $2.243\pm 0.014$ & $2.238\pm 0.015$ & $2.265\pm 0.014$ & $2.250\pm 0.014$ &  $2.266\pm 0.014$ \\
$\Omega_\chi h^2$ & $0.1192\pm 0.0010$ & $0.1199\pm 0.0012$ & $0.1169\pm 0.0011$ & $0.1184\pm 0.0009$ & $0.1170\pm 0.0008$\\
$100\theta_s$ & $1.0419\pm 0.0003$ & $1.0419\pm 0.0003$ & $1.0419\pm 0.0003$ & $1.0419\pm 0.0003$  & $1.0420\pm 0.0003$ \\
$10^{-2}\ln\lambda$ & $<-2.2$ & $<-2.2$ & $<-2.2$ & $<-2.2$ & $<-2.2$ \\ 
$10^{-2}\ln\tilde\lambda$ & $<-2.33$ & $<-2.33$ & $<-2.33$ & $<-2.33$ & $<-2.33$ \\ 
$\ln m_\chi$ & $>-43.6$ & $>-43.64$ & $>-43.58$ & $>-43.81$ &  $>-43.72$  \\
$H_0$ & $67.73_{-0.52}^{+1.80}$ & $67.40_{-0.08}^{+2.40}$ & $68.84_{-0.24}^{+2.10}$ & $68.10_{-0.48}^{+1.80}$ & $68.81_{-0.67}^{+1.60}$ \\
$\Omega_{\rm m}$ & $0.3102_{-0.0092}^{+0.0077}$ & $0.315_{-0.012}^{+0.013}$ & $0.296_{-0.008}^{+0.012}$ & $0.3052_{-0.0079}^{+0.0086}$ & $0.2963_{-0.0094}^{+0.0062}$ \\
$\Omega_\phi$ & $0.6897_{-0.0007}^{+0.0230}$ & $0.685_{-0.003}^{+0.031}$ & $0.704_{-0.000}^{+0.025}$ & $0.6948_{-0.0008}^{+0.0229}$ & $0.7036_{-0.0040}^{+0.0200}$\\
$\sigma_8$ & $0.8086_{-0.0010}^{+0.0350}$ & $0.8103_{-0.0021}^{+0.0250}$ & $0.803_{-0.011}^{+0.040}$ & $0.8061_{-0.0034}^{+0.0329}$ & $0.8043_{-0.0006}^{+0.0280}$ \\
$S_8$ & $0.822_{-0.032}^{+0.014}$ & $0.829_{-0.028}^{+0.016}$ & $0.7975_{-0.0250}^{+0.0180}$ & $0.813_{-0.031}^{+0.028}$ & $0.7993_{-0.0140}^{+0.0410}$ \\
\hline
$\Delta\chi^2_{\rm min}$ & $0.0$  & $0.0$  & $-1.0$  & $+1.0$  & $-1.0$ \\
\hline\hline 
\end{tabu}}}
\caption{Constraints on some of the cosmological parameters of our model. The values are quoted at 68\% CL intervals, unless an upper or lower bounds are shown, in which case it is the 95\% CL interval. {The lowermost row shows $\Delta\chi^2_{\rm min}=\chi^2_{\rm min,iDMDE}-\chi^2_{\text{min,}\Lambda\text{CDM}}$, where iDMDE stands for our interacting dark matter-dark energy model.} }
\label{tab2}
\end{table}

In table~\ref{tab2} we show the constraints on some of the parameters from the different combinations of data sets considered. We find that for the data sets Planck+Lensing and Planck+BAO, our model is consistent with $\Lambda$CDM with a slight shift in the central values of the parameters (still within the error bars). With the addition of the Pantheon and SH0ES as well as the WiggleZ data sets, the central values for most of the parameters have shifted. We notice that in the presence of BAO the uncertainties in the parameters are reduced, given how well the BAO observables are measured. The  inferred values of $H_0$ and $S_8$ are intriguing as they potentially can address the tension between the values obtained from Planck measurements based on $\Lambda$CDM and those measured directly, known as local measurements. For the Hubble parameter, the tension is the most serious with a significance reaching more than $5\sigma$. The Planck measurements indicate a value $H_0^{\rm Pl}=(67.4\pm 0.5)$ km/s/Mpc~\cite{Planck:2018vyg} while the most recent direct measurement from the SH0ES collaboration using Cepheids-calibrated supernovae gives $H_0^{\rm R22}=(73.04\pm 1.04)$ km/s/Mpc~\cite{Riess:2021jrx} (both at 68\% CL). Based on the fourth column of table~\ref{tab2}, including the Pantheon+SH0ES data set, we obtain $H_0=68.84_{-0.24}^{+2.10}$ km/s/Mpc which shows a movement of the central value toward the R22 measurement. Despite not being a resolution to the tension, the obtained $H_0$ is now $\sim 2.7\sigma$ away from the R22 measurement. {For the entire data sets combined together, the last column of table~\ref{tab2} shows the central value of $H_0$ barely change compared to the Planck+Pantheon+SH0ES data set (fourth column). However, the error bars are now reduced and the tension with the R22 measurement increases to $\sim 2.8\sigma$.} The alleviation of the Hubble tension is definitely slightly artificial because of the large error bars on our $H_0$ value, but one cannot ignore the fact that the central value of $H_0$ has increased relative to $\Lambda$CDM. The mean value slightly decreases and the constraints are tightened with the inclusion of the BAO and WiggleZ data sets in the last column of table~\ref{tab2}. The presence of large error bars in several of our model parameters is attributed to the presence of additional parameters in our model which makes it harder to constrain, as opposed to $\Lambda$CDM. {Before moving on to discuss $S_8$, we note that recently the Dark Energy Spectroscopic Instrument (DESI)~\cite{DESI:2024mwx} released their results on BAO measurements in galaxy, quasar and Lyman-$\alpha$ forest tracers from the first year of observations. The collaboration determined the value of the Hubble parameter in light of the new data combining DESI BAO and BBN sets to find $H_0=(68.53\pm 0.80)$ km/s/Mpc, which, even though higher than the Planck preferred value, is still at $\sim 3.4\sigma$ tension with SH0ES. It would interesting to check our model against the new DESI results in a future work.}

Planck measurements indicate that the matter density fraction is $\Omega_{\rm m}=0.315\pm 0.007$ which agrees well with the value predicted by our model using the first two data sets. This value decreases after adding the Pantheon and SH0ES in our analysis. This can be easily understood, since requiring a larger $H_0$ while having $\theta_s$ fixed means that the DM density and therefore $\Omega_{\rm m}$ must decrease to accommodate this change. Notice also that with these data sets, our model favors a universe with a smaller DM relic density, $\Omega_\chi h^2$. Turning our attention to $S_8$, we observe a similar trend: $S_8$ decreases for the last two data sets owing to a smaller matter density fraction. The Planck analysis gives $S_8^{\rm Pl}=0.834\pm 0.016$ which is larger than what is obtained from the latest cosmic shear data of KiDS-1000 and DES-Y3, giving: $S_8^{\rm KiDS}=0.759_{-0.021}^{+0.024}$~\cite{KiDS:2020suj} and $S_8^{\rm DES}=0.759_{-0.023}^{+0.025}$~\cite{DES:2021bvc,DES:2021vln}. Again, we see a consistency between our model predictions and $\Lambda$CDM for the first two sets, but the third set renders $S_8=0.7975_{-0.0250}^{+0.0180}$, a value consistent with both KiDS and DES, thus resolving the $\sim 3\sigma$ tension that $S_8$ has with the Standard Model. {However, when all the data sets are combined (fourth column in table~\ref{tab2}), the central value of $S_8$ moves up and now our value is discrepant at the $\sim 1.1\sigma$ level with the DES and KiDS results which may well be within experimental and theoretical uncertainties. Note that another model has also shown promise in resolving this tension~\cite{Poulin:2022sgp} and is based on including a drag term between DM and DE at the level of the velocity divergence equations. 

The last row of table~\ref{tab2} shows $\Delta\chi^2_{\rm min}$ representing the goodness-of-the-fit, comparing our model to $\Lambda$CDM. One can see that the first two data sets show no difference between our model and $\Lambda$CDM, whereas the third data set and the combination of all data show that our model better fits the data, albeit very slightly. 

For the additional model parameters, we set upper limits on the DM self-interaction strength $\lambda$ and the DM-DE interaction strength $\tilde\lambda$ at 95\% CL
\begin{align}
    &\lambda < 2.85\times 10^{-96}\,, \nonumber \\
    &\tilde\lambda < 6.45\times 10^{-102}\,,
\end{align}
and we set a lower limit on the mass of an ultralight DM scalar field constituting all of the DM density today
\begin{equation}
    m_\chi > 1.03\times 10^{-19}\,\text{eV}.
\end{equation}
This limit is relaxed if the ultralight scalar field constitutes a fraction of the DM density.
 The above lower limit indicates that if an ultralight scalar DM is to make up the entire DM density, then it will be hard to distinguish it from the standard CDM scenario. The reason is that the matter power spectrum of such a field will only deviate from $\Lambda$CDM at very small scales, for which non-linear effects become important. Furthermore, the temperature power spectrum of the scalar DM will track almost exactly the one for CDM as we have seen in earlier analysis. At the background level, heavier scalar DM begins to dilute as CDM very early on and so its effects on matter-radiation equality or recombination become negligible. 

\section{Conclusions\label{sec:conc}} 

 The aim of this work is to study the evolution of interacting dark matter and dark energy fields
     from early times to late times and try to fit cosmological data within self-consistent Lagrangian
     field theory involving two ultralight fields, i.e.,  a real scalar DM field $\chi$ with self-interaction and a quintessence field $\phi$ as DE. We allow for interaction between the two 
   DM and DE fields which leads to source terms $Q_{\chi}$ and $Q_{\phi}$ in the continuity equations for $\chi$ and $\phi$ which are self-consistently determined. This is in contrast to the frequently used procedure where one assumes the following set of equations 
    \begin{equation}
 {\cal D}_\alpha T_\phi^{\alpha \beta} = J_\phi^\beta~~~\text{and}~~~
  {\cal D}_\alpha T_\chi^{\alpha \beta} = J_\chi^\beta, 
 \end{equation} 
where ${\cal D}_{\alpha}(T_{\phi}^{\alpha\beta}+ T_{\chi}^{\alpha\beta})=0$ is a constraint which is introduced ad hoc in concordance models 
and does not necessarily arise from any fundamental Lagrangian. In contrast, our
approach is purely field theory where energy-momentum conservation is a consequence
of its internal consistency.
In fact
we find that $Q_\phi=J^0_\phi$ and $Q_\chi=J^0_{\chi}$ hugely deviate from the two-fluid 
assumption of $Q_\phi/Q_\chi=-1$ as can be clearly seen from the bottom right panels of Figs.~\ref{fig1},~\ref{fig4} and~\ref{fig7}.

Within the above framework, we have carried out an analysis of background and linear perturbations, in which the latter is performed in the 
 general gauge and then cast in the synchronous gauge for numerical analysis. 
The analysis of the background and perturbation equations
include self-interactions of dark matter as well as interactions between dark matter and dark energy.
  Thus one of the aims of the analysis is to study the effects of DM-DE interactions and DM self-interaction on the 
  growth of density perturbations in time. We work in the generalized dark matter scheme where we derive the sound speed of perturbations in the DM fluid and the DM equation of state to show their dependence on DM self-interaction and on DM-DE interaction. We then confront the model parameters with  the available cosmological data from Planck, BAO, Pantheon, SH0ES and WiggleZ. Using a Bayesian inference tool, we derived constraints on the parameters showing that the data favors some level of DM-DE interaction as well as DM self-interaction. Our results also show that the model discussed in this work does
 alleviate the $H_0$ tension in some data sets while resolving the $S_8$ tension.  
 Thus in summary,  the analysis allows for mild interaction  between
 the DM-DE fields and also of self-interaction while maintaining a quality of fit to all of the cosmological data comparable to that of the $\Lambda$CDM model. The analysis provides encouraging signs for possible improvements in fits to the cosmological data with more general DE and DM Lagrangian structure.

\vspace{1cm}

{\bf Acknowledgments:} One of the authors  (AA) would like to thank the University of Muenster for allocating computing resources on the Palma cluster. The research of PN was supported in part by the NSF Grant PHY-2209903.

\appendix

\section{Perturbation equations before the onset of rapid oscillations\label{app:A}}

In this section we give the form of the perturbation equations in both the conformal (newtonian) gauge and the synchronous gauge after imposing the criteria of Eqs.~(\ref{sync}) and~(\ref{conf}). These equations describe the evolution of DM and DE perturbation fields before the onset of the DM rapid oscillations about the minimum of its potential.  

The evolution of the DM and DE field perturbations are tracked by solving the Klein-Gordon equations. In the conformal gauge, the equations are given by
\begin{align}
&\phi_1^{\prime\prime}+2\mathcal{H}\phi_1^\prime+(k^2+a^2\Bar{V}_{,\phi\phi})\phi_1+a^2\Bar{V}_{,\phi\chi}\chi_1+2a^2\Bar{V}_{,\phi}\Psi-4\Psi^\prime\phi_0^\prime=0, \\
&\chi_1^{\prime\prime}+2\mathcal{H}\chi_1^\prime+(k^2+a^2\Bar{V}_{,\chi\chi})\chi_1+a^2\Bar{V}_{,\chi\phi}\phi_1+2a^2\Bar{V}_{,\chi}\Psi-4\Psi^\prime\chi_0^\prime=0\,,
\end{align}
while in the synchronous gauge they are
\begin{align}
&\phi_1^{\prime\prime}+2\mathcal{H}\phi_1^\prime+(k^2+a^2\Bar{V}_{,\phi\phi})\phi_1+a^2\Bar{V}_{,\phi\chi}\chi_1+\frac{1}{2}h^\prime\phi_0^\prime=0, \\
&\chi_1^{\prime\prime}+2\mathcal{H}\chi_1^\prime+(k^2+a^2\Bar{V}_{,\chi\chi})\chi_1+a^2\Bar{V}_{,\chi\phi}\phi_1+\frac{1}{2}h^\prime\phi_0^\prime=0\,.
\end{align}
The obtained values of $\phi_1$ and $\chi_1$ are then used to calculate the density and pressure perturbations of the two fields in the conformal gauge using
\begin{align}
    \delta\rho_\phi&=\frac{1}{a^2}\phi_0^\prime\phi_1^\prime-\frac{1}{a^2}\phi_0^{\prime 2}\Psi+(\Bar{V}_2+\Bar{V}_3)_{,\phi}\phi_1+\Bar{V}_{3,\chi}\chi_1, \\
    \delta p_\phi&=\frac{1}{a^2}\phi_0^\prime\phi_1^\prime-\frac{1}{a^2}\phi_0^{\prime 2}\Psi-(\Bar{V}_2+\Bar{V}_3)_{,\phi}\phi_1-\Bar{V}_{3,\chi}\chi_1, \\
    \delta\rho_\chi&=\frac{1}{a^2}\chi_0^\prime\chi_1^\prime-\frac{1}{a^2}\chi_0^{\prime 2}\Psi+(\Bar{V}_1+\Bar{V}_3)_{,\chi}\chi_1+\Bar{V}_{3,\phi}\phi_1, \\
    \delta p_\chi&=\frac{1}{a^2}\chi_0^\prime\chi_1^\prime-\frac{1}{a^2}\chi_0^{\prime 2}\Psi-(\Bar{V}_1+\Bar{V}_3)_{,\chi}\chi_1-\Bar{V}_{3,\phi}\phi_1,
\end{align}
and in the synchronous gauge using
\begin{align}
    \delta\rho_\phi&=\frac{1}{a^2}\phi_0^\prime\phi_1^\prime+(\Bar{V}_2+\Bar{V}_3)_{,\phi}\phi_1+\Bar{V}_{3,\chi}\chi_1, \\
    \delta p_\phi&=\frac{1}{a^2}\phi_0^\prime\phi_1^\prime-(\Bar{V}_2+\Bar{V}_3)_{,\phi}\phi_1-\Bar{V}_{3,\chi}\chi_1, \\
    \delta\rho_\chi&=\frac{1}{a^2}\chi_0^\prime\chi_1^\prime+(\Bar{V}_1+\Bar{V}_3)_{,\chi}\chi_1+\Bar{V}_{3,\phi}\phi_1, \\
    \delta p_\chi&=\frac{1}{a^2}\chi_0^\prime\chi_1^\prime-(\Bar{V}_1+\Bar{V}_3)_{,\chi}\chi_1-\Bar{V}_{3,\phi}\phi_1.
\end{align}
The background fields $\chi_0$ and $\phi_0$ are also needed in the evaluation of the perturbations. They are calculated using the Klein-Gordon equations, Eqs.~(\ref{KGc0}) and~(\ref{KGp0}).

\section{Perturbation equations after the onset of rapid oscillations\label{app:B}}

In this section we give the form of the perturbation equations in both the conformal (newtonian) gauge and the synchronous gauge after imposing the criteria of Eqs.~(\ref{sync}) and~(\ref{conf}). These equations describe the evolution of DM and DE perturbation fields after the onset of the DM rapid oscillations about the minimum of its potential.  

We work in the generalized dark matter scheme and turn the perturbation equations from the previous section to differential equations in $\delta$ (density contrast) and $\Theta$ (velocity divergence). To do so we need to calculate the sound speed, the adiabatic sound speed and the equation of state of the fields. First, we begin by showing the equations for the density contrast for the fields $\chi$ and $\phi$ in the conformal gauge:
\begin{align}
\delta_\phi^\prime&=\left[3\mathcal{H}(w_\phi-c_\phi^2)-\frac{Q_\phi}{\rho_\phi}\right]\delta_\phi+\frac{3\mathcal{H}Q_\phi}{\rho_\phi(1+w_\phi)}(c_\phi^2-c^2_{\phi_{\rm ad}})\frac{\Theta_\phi}{k}-9\mathcal{H}^2(c_\phi^2-c^2_{\phi_{\rm ad}})\frac{\Theta_\phi}{k}-\Theta_\phi k \nonumber \\
&+\frac{a^2}{k}\frac{\rho_\chi}{\rho_\phi}\Bar{V}_{3,\chi\chi}\Theta_\chi+\frac{1}{\rho_\phi}\Bar{V}_{3,\phi\chi}\chi_0^\prime\phi_1+\frac{1}{\rho_\phi}\bar{V}_{3,\chi}\chi_1^\prime+3\Psi^\prime (1+w_\phi),
\label{dphiappc}
\end{align}
and
\begin{align}
\delta_\chi^\prime&=\left[3\mathcal{H}(w_\chi-c_\chi^2)-\frac{Q_\chi}{\rho_\chi}\right]\delta_\chi+\frac{3\mathcal{H}Q_\chi}{\rho_\chi(1+w_\chi)}(c_\chi^2-c^2_{\chi_{\rm ad}})\frac{\Theta_\chi}{k}-9\mathcal{H}^2(c_\chi^2-c^2_{\chi_{\rm ad}})\frac{\Theta_\chi}{k}-\Theta_\chi k \nonumber \\
&+\frac{a^2}{k}\frac{\rho_\phi}{\rho_\chi}\Bar{V}_{3,\phi\phi}\Theta_\phi+\frac{1}{\rho_\chi}\Bar{V}_{3,\chi\phi}\phi_0^\prime\chi_1+\frac{1}{\rho_\chi}\bar{V}_{3,\phi}\phi_1^\prime+3\Psi^\prime(1+w_\chi),
\label{dchiappc}
\end{align}
while in the synchronous gauge, the equations become
\begin{align}
\delta_\phi^\prime&=\left[3\mathcal{H}(w_\phi-c_\phi^2)-\frac{Q_\phi}{\rho_\phi}\right]\delta_\phi+\frac{3\mathcal{H}Q_\phi}{\rho_\phi(1+w_\phi)}(c_\phi^2-c^2_{\phi_{\rm ad}})\frac{\Theta_\phi}{k}-9\mathcal{H}^2(c_\phi^2-c^2_{\phi_{\rm ad}})\frac{\Theta_\phi}{k}-\Theta_\phi k \nonumber \\
&+\frac{a^2}{k}\frac{\rho_\chi}{\rho_\phi}\Bar{V}_{3,\chi\chi}\Theta_\chi+\frac{1}{\rho_\phi}\Bar{V}_{3,\phi\chi}\chi_0^\prime\phi_1+\frac{1}{\rho_\phi}\bar{V}_{3,\chi}\chi_1^\prime-\frac{1}{2}h^\prime (1+w_\phi),
\label{dphiapps}
\end{align}
and
\begin{align}
\delta_\chi^\prime&=\left[3\mathcal{H}(w_\chi-c_\chi^2)-\frac{Q_\chi}{\rho_\chi}\right]\delta_\chi+\frac{3\mathcal{H}Q_\chi}{\rho_\chi(1+w_\chi)}(c_\chi^2-c^2_{\chi_{\rm ad}})\frac{\Theta_\chi}{k}-9\mathcal{H}^2(c_\chi^2-c^2_{\chi_{\rm ad}})\frac{\Theta_\chi}{k}-\Theta_\chi k \nonumber \\
&+\frac{a^2}{k}\frac{\rho_\phi}{\rho_\chi}\Bar{V}_{3,\phi\phi}\Theta_\phi+\frac{1}{\rho_\chi}\Bar{V}_{3,\chi\phi}\phi_0^\prime\chi_1+\frac{1}{\rho_\chi}\bar{V}_{3,\phi}\phi_1^\prime-\frac{1}{2}h^\prime(1+w_\chi).
\label{dchiapps}
\end{align}
The velocity divergences of the fields in the conformal gauge are given by
\begin{align}
\Theta^\prime_\phi&=(3c_\phi^2-1)\mathcal{H}\Theta_\phi+k\delta_\phi c_\phi^2+3\mathcal{H}(w_\phi-c^2_{\phi_{\rm ad}})\Theta_\phi \nonumber \\
&~~~-\frac{Q_\phi}{\rho_\phi}\left(1+\frac{c_\phi^2-c^2_{\phi_{\rm ad}}}{1+w_\phi}\right)\Theta_\phi+\frac{k}{\rho_\phi}\Bar{V}_{3,\chi}\chi_1+k(1+w_\phi)\Psi,
\label{thetaphiappc}
\end{align}
and
\begin{align}
\Theta^\prime_\chi&=(3c_\chi^2-1)\mathcal{H}\Theta_\chi+k\delta_\chi c_\chi^2+3\mathcal{H}(w_\chi-c^2_{\chi_{\rm ad}})\Theta_\chi \nonumber \\
&~~~-\frac{Q_\chi}{\rho_\chi}\left(1+\frac{c_\chi^2-c^2_{\chi_{\rm ad}}}{1+w_\chi}\right)\Theta_\chi+\frac{k}{\rho_\chi}\Bar{V}_{3,\phi}\phi_1+k(1+w_\chi)\Psi\,.
\label{thetachiappc}
\end{align}
Whereas, in the synchronous gauge, the equations take the form
\begin{align}
\Theta^\prime_\phi&=(3c_\phi^2-1)\mathcal{H}\Theta_\phi+k\delta_\phi c_\phi^2+3\mathcal{H}(w_\phi-c^2_{\phi_{\rm ad}})\Theta_\phi \nonumber \\
&~~~-\frac{Q_\phi}{\rho_\phi}\left(1+\frac{c_\phi^2-c^2_{\phi_{\rm ad}}}{1+w_\phi}\right)\Theta_\phi+\frac{k}{\rho_\phi}\Bar{V}_{3,\chi}\chi_1,
\label{thetaphiapps}
\end{align}
and
\begin{align}
\Theta^\prime_\chi&=(3c_\chi^2-1)\mathcal{H}\Theta_\chi+k\delta_\chi c_\chi^2+3\mathcal{H}(w_\chi-c^2_{\chi_{\rm ad}})\Theta_\chi \nonumber \\
&~~~-\frac{Q_\chi}{\rho_\chi}\left(1+\frac{c_\chi^2-c^2_{\chi_{\rm ad}}}{1+w_\chi}\right)\Theta_\chi+\frac{k}{\rho_\chi}\Bar{V}_{3,\phi}\phi_1.
\label{thetachiapps}
\end{align}
The speed of sound and the adiabatic sound speed of species $i$ are given by
\begin{align}
    &c^2_{s i}=\frac{\delta p_i}{\delta\rho_i} \\
    &c^2_{i_{\rm ad}}\equiv\frac{p^\prime_i}{\rho^\prime_i}=w_i-\frac{w^\prime_i \rho_i}{3\mathcal{H}(1+w_i)\rho_i-Q_i}.
\end{align}
Working in the gauge comoving with the DM fluid, we arrive at the sound speed given by Eq.~(\ref{sound}). Furthermore, the additional model-dependent terms appearing in Eqs.~(\ref{dchiappc}),~(\ref{dchiapps}),~(\ref{thetachiappc}) and~(\ref{thetachiapps}) having the following averages
\begin{align}
    \Big\langle\frac{Q_\chi}{\rho_\chi}\Big\rangle&=\tilde{\lambda}\phi_0\phi_0^\prime\left(\frac{1-3w_\chi}{m_\chi^2+\tilde{\lambda}\phi_0^2}\right), \\
    \Big\langle\frac{V_{3,\phi\phi}}{\rho_\chi}\Big\rangle&=\frac{\tilde{\lambda}(1-3w_\chi)}{m_\chi^2+\tilde{\lambda}\phi_0^2}, \\
    \Big\langle\frac{V_{3,\phi\chi}\phi_0^\prime\chi_1}{\rho_\chi}\Big\rangle&=-4\tilde{\lambda}\phi_0\phi_0^\prime\left(\frac{1-3w_\chi}{m_\chi^2+\tilde{\lambda}\phi_0^2}\right)\left(\frac{a^2 m_\chi^2+\tilde{\lambda}a^2(\phi_0^2+\phi_0\phi_1)}{k^2+\tilde{\lambda}a^2\phi_0^2}\right)\Psi, \\
    \Big\langle\frac{V_{3,\phi}\phi_1^\prime}{\rho_\chi}\Big\rangle&=\tilde{\lambda}\phi_0\phi_1^\prime\left(\frac{1-3w_\chi}{m_\chi^2+\tilde{\lambda}\phi_0^2}\right), \\
    \Big\langle\frac{V_{3,\phi}\phi_1}{\rho_\chi}\Big\rangle&=\tilde{\lambda}\phi_0\phi_1\left(\frac{1-3w_\chi}{m_\chi^2+\tilde{\lambda}\phi_0^2}\right).
\end{align}

\end{document}